\documentclass{achemso}
\usepackage{amsmath}
\usepackage{amssymb}
\usepackage{amsfonts}
\usepackage{multirow}
\usepackage{threeparttable}
\usepackage{booktabs}
\usepackage{bm}
\usepackage{dcolumn}
\usepackage[T1]{fontenc}
\usepackage[version=3]{mhchem}
\usepackage{subcaption}
\usepackage{caption}
\usepackage{graphicx}
\usepackage{xcolor}
\usepackage{accents}
\usepackage[normalem]{ulem}
\usepackage[colorlinks,linkcolor=blue,urlcolor=blue,filecolor=blue,citecolor=blue]{hyperref}
\setkeys{acs}{articletitle=true}
\SectionNumbersOn

\newlength{\dhatheight}
\newcommand{\hathat}[1]{%
    \settoheight{\dhatheight}{\ensuremath{\hat{#1}}}%
    \addtolength{\dhatheight}{-0.35ex}%
    \hat{\vphantom{\rule{1pt}{\dhatheight}}%
    \smash{\hat{#1}}}}
\graphicspath{{figure/}}
\makeatother

\author{Yu Zhang}
\affiliation{School of Chemistry and Biological Engineering,
University of Science and Technology Beijing, Beijing 100083, China}
\author{Junzi Liu}
\email{jliu413@ustb.edu.cn}
\affiliation{School of Chemistry and Biological Engineering,
University of Science and Technology Beijing, Beijing 100083, China}

\title{Unitary Coupled-Cluster based Self-Consistent Electron Propagator Theory
for Electron-Detached and Electron-Attached States: A Quadratic Unitary Coupled-Cluster
Singles and Doubles Method and Benchmark Calculations}

\begin{document}

\begin{tocentry}
  \centering
  \includegraphics[width=0.9\textwidth]{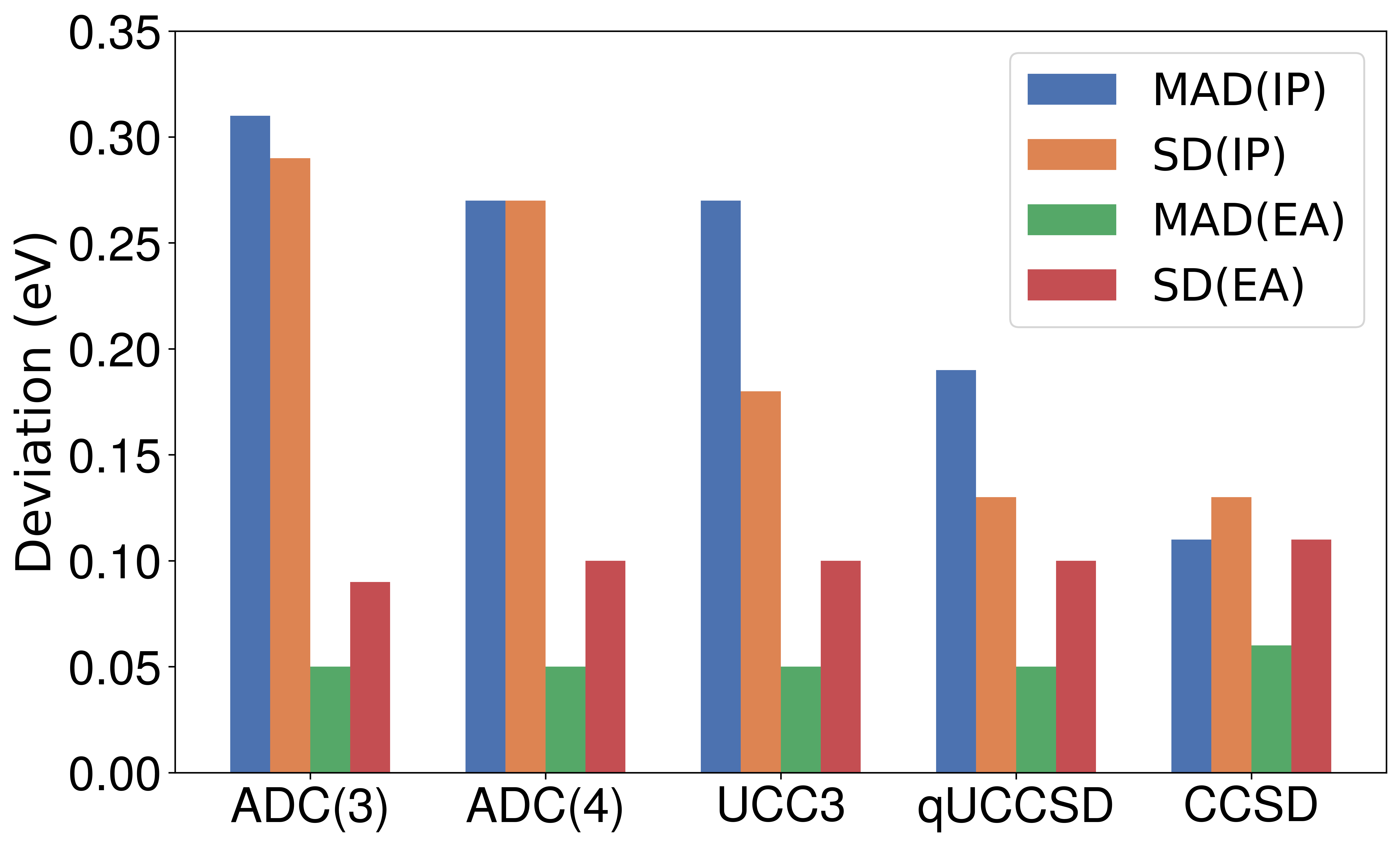}
\end{tocentry}

\begin{abstract}
A unitary coupled-cluster (UCC)-based self-consistent electron propagator theory
(EPT) is proposed for the description of electron-detached and
electron-attached states. Two practical schemes, termed IP/EA-UCC3 and
IP/EA-qUCCSD, are developed and implemented within the UCC singles and doubles
(UCCSD) framework using the perturbative and commutator-based truncation
{strategies} for the similarity-transformed Hamiltonian $\bar{H}$. The numerical
performance of these UCC-based EPT methods is evaluated 
{primarily using} full configuration interaction (FCI) reference data and
compared with established approaches, including IP/EA-ADC(3), IP/EA-ADC(4) and
IP/EA-EOM-CCSD. Benchmark calculations demonstrate that IP-qUCCSD achieves
the highest overall accuracy among Hermitian ionized-state methods for
one-hole (1h)-dominated IPs of closed-shell systems, with a mean absolute
deviation (MAD) of 0.19 eV and standard deviation (SD) of 0.13 eV.
Remarkably, despite the absence of triple-excitation contributions, IP-qUCCSD
outperforms the higher-order ADC(4) method. For one-particle (1p)-dominated
EA calculations, all tested methods exhibit comparable accuracy.
\end{abstract}

\section{Introduction}
Propagator theory provides a general and powerful framework in quantum chemistry
for describing the dynamical and response properties of many-electron systems
through the formalism of propagators, also known as many-body Green's functions.
\cite{Linderberg1973,*Linderberg2004,Pickup1973,Cederbaum1975,Cederbaum1977,
Oddershede1987,Goscinski1980-1,Jorgensen1981,Schirmer1982,Schirmer1983,Oddershede1984,
Kutzelnigg1989,Longo1995,Simons2005,Danovich2011,Ortiz2012,Schirmer2018}
It comprises two principal branches: the polarization propagator theory (PPT) and
electron propagator theory (EPT).
PPT characterizes neutral excitations by modeling how perturbations in the electron
density propagate through a correlated many-electron system, thereby yielding
excitation energies (EEs) and related properties.
In contrast, EPT focuses on single-particle propagation, providing direct access
to electron binding energies, including ionization potentials (IPs) and electron
affinities (EAs).
Together, PPT and EPT constitute a unified theoretical formalism for describing
both charged and neutral excitations in molecules and {condensed phase.}
The present work focuses on EPT, while detailed developments of PPT
are presented in refs.\
\citenum{Oddershede1978,Oddershede1987,Bak2000,Faleev2004,van_Schilfgaarde2006,
van_Setten2013,Reining2018,Geertsen1987,Kowalski2014,Vila2021,Peng2021,Salpeter1951,
Strinati1988,Blase2020,Schirmer1991,Trofimov1995,Mertins1996-I,Trofimov1999,Trofimov2002,
Schirmer2004,Starcke2009,Dreuw2015,Leitner2022,Sokolov2018,Mazin2021,Dutta2025,
Liu2018,Liu2021,Liu2022,Liu2025,Hodecker2020_UCC3,Hodecker2020,Dutta2024}.
Owing to its rigorous treatment of electron correlation and its direct
connection to optical and X-ray photoelectron spectra,\cite{Hufner2003,VanderHeide2011}
EPT-based approaches have become powerful and extensively developed tools for
electronic-structure and spectroscopic studies.

The electron propagator $G(\omega)$ (one-particle Green's function) is formally
related to the non-interacting Green's function $G^0(\omega)$ through the Dyson
equation, in which electron correlation beyond the mean-field approximation and
orbital relaxation effects are incorporated via the nonlocal, frequency-dependent
effective one-particle potential $\Sigma(\omega)$, commonly referred to as the
self-energy.\cite{Schirmer2018}
In practice, solving the Dyson equation requires suitable approximations to
$\Sigma(\omega)$, which are typically constructed using
diagrammatic perturbation theory. This leads to a variety of finite-order
perturbative expansion approaches, including
the outer valence Green's function (OVGF) method,\cite{Cederbaum1975,Cederbaum1977}
the diagonal second order (D2) method,\cite{Cederbaum1971,Doll1972,Cederbaum1973}
the diagonal third order (D3) method,\cite{Cederbaum1973}
Dyson equation-based algebraic diagrammatic construction (Dyson-ADC) methods
\cite{Schirmer1983,Vonniessen1984,Angonoa1987,Schirmer1989},
the two-particle-one-hole Tamm-Dancoff approximation (2p1h-TDA),\cite{Schirmer1978}
the partial third-order (P3) method,\cite{Ortiz1996}
the quasiparticle third-order (Q3) method,\cite{Opoku2021}
and the {linear third-order} (L3) methods,\cite{Opoku2021} along with their various renormalized
extensions and variants.\cite{Ortiz1998,Ortiz2005,Opoku2021,Opoku2024}
Collectively, these can be called Dyson-type approaches.
Within Dyson-type EPT, the electron-attachment and electron-detachment processes
are treated simultaneously through a coupled description of the ($N \pm 1$)-electron states.
On the other hand, the electron propagator can be rigorously decomposed into a
spectral representation by introducing a complete set of ($N \pm 1$)-electron
states and the exact ground-state wavefunction. This decomposition naturally
gives rise to forward and backward components of electron propagator,
corresponding to electron attachment and detachment, respectively.
Within this formulation, the approximated ground-state and the ($N \pm
1$)-electron states can be employed explicitly, thereby avoiding the need to
solve the Dyson equation; such methods are therefore often referred to as
non-Dyson (nD) approaches. In this context, several non-Dyson approaches have been
formulated and implemented, including the nD-D2, nD-D3, nD-L3,
\cite{Opoku2024_JPCA,Opoku2024} and nD-ADC methods.
\cite{Schirmer1998,Trofimov2005,Storchi2009,Banerjee2019,Banerjee2022,Banerjee2023}
In contrast to Dyson-type methods, non-Dyson approaches decouple electron
attachment and electron-detachment components of the electron propagator,
making them conceptually more transparent and computationally more efficient.
It is notable that the ADC family of methods is perhaps the most
popular EPT/PPT approach, attracting considerable interest as a Hermitian
excited-state framework that offers a favorable balance between computational
efficiency and accuracy.
Over the past decades, ADC methods for computing electron-detached and
electron-attached energies, as well as {the} related properties, have been
systematically developed to various perturbative orders through diagrammatic
techniques,\cite{Schirmer1998,Storchi2009,Dempwolff2019}
the intermediate state representation (ISR) framework,
\cite{Schirmer1991,Trofimov2005,Schirmer2001,Thiel2003,Schneider2015,
Dempwolff2020-IP-I, Dempwolff2020-IP-II,Dempwolff2021-EA,Leitner2024}
and effective Liouvillean theory.\cite{Banerjee2019,Banerjee2021}

Among the various approaches related to EPT, those based on coupled-cluster
theory -- such as the coupled-cluster Green's function (CCGF) method,
\cite{Hu1990,Nooijen1992,Nooijen1993,Nooijen1995a,Meissner1993,Bhaskaran2016,Peng2018} 
the symmetry-adapted cluster configuration interaction (SAC-CI) method,
\cite{Nakatsuji1978_JCP,Nakatsuji1978_CPL,Nakatsuji1979,Nakatsuji1981,Nakajima1997}
and the equation-of-motion coupled-cluster (EOM-CC)
\cite{Sekino1984,Geertsen1989,Stanton1993,Bartlett1994,Mertins1996-II,
Kowalski2000,Kucharski2001,Hirata2004,Kallay2004,Levchenko2004,Matthews2016}
/linear-response CC (LR-CC) theories\cite{Monkhorst1977,Mukherjee1979,Ghosh1981,
Emrich1981,Dalgaard1983,Koch_CCRF_1990,Koch_CCSDLR_1990,Christiansen1995}
-- have established themselves as state-of-the-art tools for the
accurate and reliable description of ($N \pm 1$)-electron
states.\cite{Krylov2008,Bartlett2012,Sneskov2012,Loos2020}
In particular, the IP/EA-EOM-CC approaches provide a hierarchy of {size-intensive}
methods whose accuracy can be systematically improved by including contributions from
higher excitations, ranging from singles and doubles to triples and even
quadruples.\cite{Stanton1994,Nooijen1995b,Stanton1999,Hirata2000,Musial2003,Musial2004,
Bomble2005,Gour2005,Kamiya2006,Kamiya2007}
{In addition, several simplified EOM-CC-based variants for computing IPs
and EAs have been developed.\cite{Stanton1995,Bozkaya2014,Walz2016,Ma2020,Boguslawski2021,
Galynska2024EA,Galynska2024,Behjou2025}}
However, because the similarity-transformed Hamiltonians in EOM-CC/LR-CC methods are
non-Hermitian, they may yield complex excitation energies,\cite{Thomas2021}
especially near conical intersections between electronic states of the same symmetry.
\cite{hattig2005,Kohn2007,Kjonstad2017_JCP,Kjonstad2017_JPCL,Kjonstad2019,
Kjonstad_theory_2024} This limitation has motivated the development of Hermitian 
formulation within the CC framework.\cite{Liu2018,Liu2021,Kutzelnigg1982,Hoffmann1988,
Bartlett1989,Watts1989,Kutzelnigg1991,Taube2006,Cooper2010,Evangelista2011,Walz2012,
Harsha2018,Phillips2025} 

{Besides} the methods discussed above, the self-consistent operator expansion
offers another valuable route for solving the equations of the
polarization/electron propagator theories.\cite{Goscinski1980-1,Goscinski1980-2,
Prasad1985,Datta1993}
Nearly forty years ago, {Mukherjee} and co-workers proposed a compact
formulation of self-consistent propagator theory that employs a unitary
couple-cluster (UCC) ground-state wavefunction together with a self-consistent
excitation manifold constructed from unitary-transformed excitation operator.\cite{Prasad1985}
This formulation uses the UCC ansatz to generate $N$-representable ground and
excited states, enabling a Hermitian and non-perturbative treatment of
neutral and charged excitations through the decoupling of forward and
backward propagator components. More recently, the working {equations} of
UCC-based self-consistent polarization propagator theory have been derived and
implemented on efficient computational platforms.\cite{Liu2018,Liu2021,Liu2022}
Because the similarity-transformed Hamiltonian $\bar{H}$ in UCC is non-terminating,
two practical truncation schemes have been developed: a third-order perturbative
approach (UCC3)\cite{Liu2018} and a quadratic commutator-truncated approach (qUCCSD).\cite{Liu2021}
These methods have been extensively benchmarked and consistently exhibit
systematic performance improvements from UCC3 to qUCCSD.\cite{Liu2022,Liu2025}
Notably, the strict third-order UCC-PPT method (UCC3-s) has been shown to be formally
equivalent to the strict third-order ADC method (ADC(3)-s, commonly referred to ADC(3)).\cite{Liu2018}
Subsequently, the theoretical connection between UCC-based PPT and
the intermediate state representation (ISR) formulation of ADC methods has been
established.\cite{Mertins1996-I,Hodecker2020_UCC2,Hodecker2020_UCC3} 
Building on this connection, UCC3 schemes for computing IPs and EAs have been
implemented and {compared with} their ADC counterparts.
\cite{Dempwolff2022,Hodecker2022,Hodecker2025}
Nevertheless, a rigorous and explicit derivation of the working equations for
UCC-based EPT using the self-consistent operator expansion is still lacking.
Moreover, qUCCSD approaches for IP and EA calculations have not yet {been}
implemented and systematically benchmarked. It therefore remains an open
question whether the systematic improvements observed from UCC3 to qUCCSD within
{polarization propagator} theory will extend in a similar manner to the electron propagator
framework.

In this work, we rigorously derive UCC-EPT {using}
the self-consistent operator expansion and present implementations of both the
third-order (IP/EA-UCC3) and quadratic commutator-truncated (IP/EA-qUCCSD)
approaches. By integrating UCC theory with the self-consistent propagator
framework, these methods provide a Hermitian, non-perturbative and
systematically improvable alternative to conventional IP/EA-EOM-CC approaches.
Their performance is evaluated through extensive benchmarks {and compared with}
established ADC methods and IP/EA-EOM-CCSD, enabling a direct assessment of accuracy,
robustness, and computational cost. This paper is organized as
follows: section 2 presents the theoretical derivation of working equations and
introduces the third-order and quadratic UCCSD schemes for computing IPs and
EAs; section 3 summarize the computational details; section 4 discusses 
benchmark results and comparisons {among IP/EA-EOM-CCSD, IP/EA-ADC(3), IP/EA-ADC(4), 
IP/EA-UCC3, and IP/EA-qUCCSD methods}; and concluding remarks and outlooks are given in section 5.

\section{Theory}
\subsection{Electron propagator theory}
A matrix element of the electron propagator $G_{p,r}(\omega)$ in the frequency
domain is defined as
\begin{equation}
  G_{p,r}(\omega) = \langle \Psi_\mathrm{gr}^N
  | [ \hat{a}_p, (\omega \hathat{I} + \hathat{H})^{-1} \hat{a}_r^\dagger ]_- |
  \Psi_\mathrm{gr}^N \rangle,
\end{equation}
where $\Psi_\mathrm{gr}^N$ denotes the exact ground-state wavefunction of the
$N$-electron system, and $\hat{a}_r^\dagger$ and $\hat{a}_p$ are the fermionic
creation and annihilation operators, respectively. 
The notation $[\ ]_-$ represents the anticommutator, $[A, B]_- = AB + BA$,
and is used here because the expression contains an odd number of fermionic
operators (either $\hat{a}_p$ or $\hat{a}_r^\dagger$).
{The action of the superoperator $\hathat{H}$ of the Hamiltonian $\hat{H}$ on 
an arbitrary operator $\hat{A}$ gives the commutator of $\hat{A}$ and $\hat{H}$, namely, }
$\hathat{H}\hat{A} = [\hat{A}, \hat{H}]$; while the identity superoperator $\hathat{I}$ satisfies
$\hathat{I}\hat{A} = \hat{A}$. The spectral representation of the one-particle
propagator can be decomposed into forward $G_{p,r}^+(\omega)$ and backward 
$G_{p,r}^-(\omega)$ components:
\begin{eqnarray}
  G_{p,r}(\omega) &=& G_{p,r}^+(\omega) + G_{p,r}^-(\omega) \\
  G_{p,r}^+(\omega) &=& 
  \sum_L \frac{ \langle \Psi_\mathrm{gr}^N | \hat{a}_p |\Psi_L^{N+1} \rangle 
  \langle \Psi_L^{N+1} | \hat{a}_r^\dagger |\Psi_\mathrm{gr}^N \rangle}
   {\omega - (E_L^{N+1} - E_\mathrm{gr}^N)}  \\
  G_{p,r}^-(\omega) &=& 
  \sum_K \frac{ \langle \Psi_\mathrm{gr}^N | \hat{a}_r^\dagger |\Psi_K^{N-1} \rangle 
  \langle \Psi_K^{N-1} | \hat{a}_p |\Psi_\mathrm{gr}^N \rangle}
   {\omega + (E_K^{N-1} - E_\mathrm{gr}^N)}
\end{eqnarray}
where $G^+_{p,r}(\omega)$ and $G^-_{p,r}(\omega)$ correspond to electron-attachment
and electron-detachment processes, respectively. The sets
$\{\Psi_{L}^{N+1}\}$ and $\{\Psi_{K}^{N-1}\}$ denote complete bases of
$(N+1)$ and $(N-1)$-electron states with associated energies $E_L^{N+1}$ and
$E_K^{N-1}$. The exact ground-state and excited-state wavefunctions are not
accessible in practical implementations of electron propagator theory.
Therefore, approximate ground-state wavefunctions and incomplete manifold of
excited states are employed. To decouple the forward and backward components of
the propagator, the approximate wavefunction must satisfy the so-called ``vacuum
annihilation condition'' (VAC).\cite{Goscinski1980-1,Goscinski1980-2,Prasad1985}
Self-consistent propagator theories\cite{Goscinski1980-1,Goscinski1980-2,
Goscinski1984,Prasad1985,Datta1993} enforce this decoupling by constructing
self-consistent excitation operators that satisfy the VAC.
The unitary coupled-cluster (UCC) ansatz can be directly integrated into this
self-consistent propagator framework. In recent years, UCC-based self-consistent
polarization propagator theory has been formulated, derived, and implemented.
\cite{Liu2018,Liu2021,Liu2022}. The generality of this formalism enables its
extension to a corresponding self-consistent electron propagator theory,
providing a foundation for UCC-based Hermitian approaches to $(N\pm1)$-electron states.

\subsection{UCC-based self-consistent electron propagator theory}
The basic theoretical framework of UCC has been presented in our previous work
\cite{Liu2018,Liu2021}. For a self-contained presentation, we briefly
recapitulate its key contents here. In UCC theory, the ground-state
wavefunction $\Psi_{\mathrm{gr}}$ is written as a unitary exponential operator
acting on a reference state $|\Phi_0\rangle$, typically the Hartree-Fock
determinant:
\begin{equation}
  |\Psi_\mathrm{gr}\rangle = e^{\hat{\sigma}}|\Phi_0\rangle.
\end{equation}
Unlike the conventional coupled-cluster theory, the cluster operator ${\hat{\sigma}}$
in UCC is defined as $\hat{\sigma} = \hat{T} - \hat{T}^\dagger$ and is therefore
anti-Hermitian, $\hat{\sigma}^\dagger = - \hat{\sigma}$. Consequently, the wave
operator $e^{\hat{\sigma}}$ is unitary:
\begin{equation}
  \left(e^{\hat{\sigma}}\right)^\dagger = e^{-\hat{\sigma}} = \left(e^{\hat{\sigma}} \right)^{-1}.
\end{equation}
In the UCC singles and doubles (UCCSD) model, the excitation and de-excitation
{operators} are
\begin{eqnarray}
\hat{T} &=& \sum_{ai}\sigma_i^a \{ \hat{a}_a^\dagger \hat{a}_i \} + \frac{1}{4}
  \sum_{abij} \sigma_{ij}^{ab} \{ \hat{a}_a^\dagger \hat{a}_b^\dagger \hat{a}_j
  \hat{a}_i \}, \\
\hat{T}^\dagger &=& \sum_{ai} (\sigma_i^a)^* \{ \hat{a}_i^\dagger \hat{a}_a \}
  + \frac{1}{4}\sum_{abij} (\sigma^{ab}_{ij})^* \{\hat{a}_i^\dagger
  \hat{a}_j^\dagger \hat{a}_b \hat{a}_a \},
\end{eqnarray}
where $\{i,j,\dots\}$, $\{a, b, \dots \}$ and $\{p,q,\dots\}$ denote occupied, 
virtual, and arbitrary orbitals, respectively, and $\sigma_i^a$
and $\sigma_{ij}^{ab}$ are the ground-state cluster amplitudes. The UCCSD
ground-state energy and amplitude equations are
\begin{eqnarray}
	E_\mathrm{gr} &=& \langle \Phi_0 | \bar{H} | \Phi_0 \rangle, \\
	0 &=& \langle \Phi_l | \bar{H} | \Phi_0 \rangle, 
\end{eqnarray}
where $\{\Phi_l\}$ comprises the singly and doubly excited determinants, i.e.,
$\{\Phi_i^a\}$ and $\{\Phi_{ij}^{ab}\}$.
Here $\bar{H} = e^{-\hat{\sigma}} \hat{H} e^{\hat{\sigma}}$ is the Hermitian
similarity-transformed Hamiltonian. The projected amplitude equations not only
determine the ground-state amplitudes but also ensure the block-diagonal
structure of the matrix representation of the electron propagator, as discussed
below.

In UCC-based self-consistent electron propagator theory, the self-consistent
operator manifolds for electron detachment and electron attachment are
$\{\hat{z}_I^\dagger\} \cup \{\hat{z}_I\}$ and $\{\hat{z}_A^\dagger\} \cup
\{\hat{z}_A\}$, respectively. Both manifolds are formed from unitary-transformed
excitation and de-excitation operators,
\begin{equation}
  \hat{z}^\dagger_I = e^{\hat{\sigma}} \hat{b}_I^\dagger e^{-\hat{\sigma}}, \ \ \
  \hat{z}^\dagger_A = e^{\hat{\sigma}} \hat{b}_A^\dagger e^{-\hat{\sigma}},
\end{equation}
where $\hat{b}_I^\dagger$ and $\hat{b}_A^\dagger$ denote the original
electron-detachment and electron-attachment operators, respectively. 
In the UCCSD model, we adopt the form $\{\hat{b}^\dagger_I\} = \{ \hat{a}_i \} \cup
\{\hat{a}_a^\dagger \hat{a}_j \hat{a}_i\}$ for electron detachment and
$\{\hat{b}_A^\dagger\} = \{ \hat{a}_a^\dagger \} \cup \{\hat{a}_a^\dagger
\hat{a}_b^\dagger \hat{a}_i\}$ for electron attachment.
To streamline the derivation, we introduce the binary product notation
\begin{equation}
  (X|Y) = \langle \Psi_\mathrm{gr}^N | [X, Y^\dagger]_- | \Psi_\mathrm{gr}^N\rangle,
  \label{bp_ab}
\end{equation}
so that the electron propagator can be rewritten as
\begin{equation}
  G_{p,r}(\omega) = \langle
  \Psi_\mathrm{gr}^N | [ \hat{a}_p, (\omega \hathat{I} + \hathat{H})^{-1} \hat{a}_r^\dagger ]_-
  | \Psi_\mathrm{gr}^N \rangle = (\hat{a}_p | (\omega\hathat{I}-\hathat{H})^{-1}\hat{a}_r). 
  \label{G_bp}
\end{equation}
We introduce the superoperator resolution of identity within the subspaces
spanned by the self-consistent operators (including both electron-attachment and
electron-detachment operators)\cite{Prasad1985}:
\begin{eqnarray}
  \hathat{I} &=& \sum_A |\hat{z}_A)(\hat{z}_A| + \sum_I|\hat{z}^\dagger_I)(\hat{z}^\dagger_I|, \label{ri}\\
  (X|Y) &=& \sum_A (X|\hat{z}_A)(\hat{z}_A|Y) + \sum_I(X|\hat{z}^\dagger_I)(\hat{z}^\dagger_I|Y). \label{bp_ri}
\end{eqnarray}
By inserting the resolution of identity [eq.~(\ref{ri})-(\ref{bp_ri})] into the
binary product form of $G_{p,r}(\omega)$ [eq.~(\ref{G_bp})], a matrix form of
the electron propagator is obtained,
\begin{equation}
  G(\omega)_{p,r} = M_p D^{-1}(\omega) M_r^\dagger,
\end{equation}
where $M_p$ and $M_r^\dagger$ contain both electron-attachment and
electron-detachment components,
\begin{eqnarray}
  M_{p} &=&
  \begin{pmatrix}
     (M_+)_{p,A}  &(M_-)_{p,I}
  \end{pmatrix}
  =
  \begin{pmatrix}
     (\hat{a}_p | \hat{z}_A) &(\hat{a}_p | \hat{z}_I^\dagger)
  \end{pmatrix}, \\
  M^\dagger_{r} &=&
  \begin{pmatrix}
    (M_+^\dagger)_{B,r} \\ (M_-^\dagger)_{J,r}
  \end{pmatrix}
  =
  \begin{pmatrix}
     (\hat{z}_B | \hat{a}_r) \\
     (\hat{z}_J^\dagger | \hat{a}_r)
  \end{pmatrix}.
\end{eqnarray}
The matrix $D$ has a $2\times2$ block structure,
\begin{equation}
  D = \begin{pmatrix} D_{++}  &D_{+-} \\ D_{-+}  &D_{--} \end{pmatrix},
\end{equation}
with matrix elements
\begin{eqnarray}
  (D_{++})_{AB} &=& (\hat{z}_A |(\omega\hathat{I} - \hathat{H})| \hat{z}_B), \\
  (D_{--})_{IJ} &=& (\hat{z}_I^\dagger |(\omega\hathat{I} - \hathat{H})| \hat{z}^\dagger_J), \\
  (D_{+-})_{AJ} &=& (\hat{z}_A |(\omega\hathat{I} - \hathat{H})| \hat{z}^\dagger_J), \\
  (D_{-+})_{IB} &=& (\hat{z}_I^\dagger |(\omega\hathat{I} - \hathat{H})| \hat{z}_B).
\end{eqnarray}
{Consistent} with VAC, the off-diagonal blocks simplify to
\begin{align}
  (D_{+-})_{AJ} &= \langle \Phi_0 | \hat{b}_A \hat{b}_J \bar{H} | \Phi_0 \rangle = 0, 
  \label{amp_eq1} \\
  (D_{-+})_{IB} &= -\langle \Phi_0 | \bar{H} \hat{b}_B^\dagger
  \hat{b}_I^\dagger | \Phi_0 \rangle = 0.  \label{amp_eq2}
\end{align}
A product of electron-attachment and electron-detachment operators
($\hat{b}_A\hat{b}_J$ and $\hat{b}_B^\dagger \hat{b}_I^\dagger$) constitutes
a conventional excitation operator; hence $\hat{b}_B^\dagger \hat{b}_I^\dagger | \Phi_0 \rangle$ 
and $\langle \Phi_0 | \hat{b}_A \hat{b}_J$ correspond to excited determinants
and their Hermitian conjugates. Therefore, eq.~\eqref{amp_eq1}-\eqref{amp_eq2}
are precisely the projected UCC amplitude equations. It follows that
$D$ is block-diagonal and the forward and backward components of electron
propagator decouple:
\begin{equation}
  D = \begin{pmatrix} D_{++} & 0 \\ 0  & D_{--} \end{pmatrix}. 
\end{equation}
The diagonal blocks $D_{++}$ and $D_{--}$ comprise matrix elements of $\bar{H}$:
\begin{eqnarray}
  & & (D_{++})_{AB} = \delta_{AB}\omega - (\bar{H})_{AB}, \hspace{1em}
  (\bar{H})_{AB} = \langle\Phi_0 | \hat{b}_A\bar{H}\hat{b}^\dagger_B | \Phi_0\rangle, \\
  & & (D_{--})_{IJ} = \delta_{IJ}\omega - (\bar{H})_{IJ}, \hspace{1.9em}
  (\bar{H})_{IJ} = \langle\Phi_0 | \hat{b}_I\bar{H}\hat{b}_J^\dagger | \Phi_0\rangle.
\end{eqnarray}
The energies of electron-detached and electron-attached states are obtained
from the poles of the corresponding electron propagator components, leading to
eigenvalue equations for the transformed Hamiltonian $\bar{H}$ within the
spaces spanned by $\{b_I^\dagger \}$ and $\{b_A^\dagger \}$:
\begin{equation}
  \sum_J(\bar{H})_{IJ} C_{JK} = E_K C_{IK},\quad
  \sum_B(\bar{H})_{AB} C_{BL} = E_L C_{AL}.
  \label{seq_eq}
\end{equation}
The corresponding wavefunctions are
\begin{equation}
  \Psi_K^{\mathrm{IP}} = \sum_I C_{IK} e^{\hat{\sigma}} \hat{b}_I^\dagger |
  \Phi_0\rangle, \quad
  \Psi_L^{\mathrm{EA}} = \sum_A C_{AL} e^{\hat{\sigma}} \hat{b}_A^\dagger |
  \Phi_0\rangle.
\end{equation}
These eigenvalue equations can also be derived within the intermediate-state
representations by employing $\hat{z}_I^\dagger |\Psi_\mathrm{gr}\rangle =
e^{\hat{\sigma}}b_I^\dagger | \Phi_0 \rangle$ and
$\hat{z}_A^\dagger|\Psi_{\mathrm{gr}}\rangle = e^{\hat{\sigma}}b_A^\dagger | \Phi_0 \rangle$
as the basis for representing the Hamiltonian.\cite{Mertins1996-I}
Within this framework, several related methods have been developed, including
IP/EA-ADC(3) and IP/EA-UCC3.
\cite{Dempwolff2020-IP-I,Dempwolff2021-EA,Hodecker2022,Hodecker2025}

It is important to emphasize that when a complete set of self-consistent
operator manifolds is employed, the decoupling between the forward and backward
components of electron propagator is exact. The truncation of the excitation
manifolds introduces approximations. For example, the UCCSD amplitude equations
guarantee the vanishing of single and double excitations/de-excitations in
$\bar{H}$, yielding exact decoupling between single excitation in the forward
propagator and single de-excitations in the backward propagator. By contrast,
the decoupling between single excitations in $G^+_{p,r}$ and double
de-excitations in $G^-_{p,r}$, as well as between double excitations in
$G^+_{p,r}$ and single/double de-excitations in $G^-_{p,r}$, remains approximate
because the triple and quadruple excitation/de-excitation terms in $\bar{H}$ do
not vanish. Consequently, the eigenvalue equations [eq.~\eqref{seq_eq}] neglect
these residual couplings. Nevertheless, since the one-particle electron
propagator theory is primarily used for describing $(N\pm1)$-electron states
with dominant single-particle or single-hole character, the exact decoupling of
single excitations and single de-excitations within the UCCSD-based electron
propagator framework is particularly attractive and useful feature.

Within UCCSD model, the components of $\bar{H}$ relevant to the ground-state
energy $E_\mathrm{gr}$, the amplitude equations ($\bar{H}_{ai}$ and
$\bar{H}_{ab,ij}$), and the IP/EA eigenvalue equations can be summarized as
\begin{eqnarray}
  \bar{H} &=&
  E_{\mathrm{gr}} + \left( \left(\bar{H}_{ai}\{\hat{a}^\dagger_a \hat{a}_i\} +
  \frac{1}{4}\bar{H}_{ab,ij}\{\hat{a}^\dagger_a \hat{a}^\dagger_b \hat{a}_j
  \hat{a}_i\} \right) + h.c. \right) +
  \left( \bar{H}_{ij} \{ \hat{a}_i^\dagger \hat{a}_j \} +
  \bar{H}_{ab} \{ \hat{a}_a^\dagger \hat{a}_b \} + \right. \nonumber \\
  & & \left.
  \frac{1}{4}\bar{H}_{ij,kl} \{ \hat{a}^\dagger_i \hat{a}^\dagger_j \hat{a}_l \hat{a}_k \} +
  \frac{1}{4}\bar{H}_{ab,cd} \{ \hat{a}^\dagger_a \hat{a}^\dagger_b \hat{a}_d \hat{a}_c \} +
  \bar{H}_{ia,bj} \{ \hat{a}^\dagger_i \hat{a}^\dagger_a \hat{a}_j \hat{a}_b \} \right) +
  \left( \left(\frac{1}{2}\bar{H}_{ij,ka}\{\hat{a}_i^\dagger \hat{a}_j^\dagger \hat{a}_a
  \hat{a}_k\} \right. \right.\nonumber \\
  & & \left. \left.
  +\frac{1}{2}\bar{H}_{ab,ci}\{\hat{a}_a^\dagger \hat{a}_b^\dagger \hat{a}_i
  \hat{a}_c \}\right) + h.c. \right).
\end{eqnarray}
The eigenvalue equations for electron-detached (ionized) states can be
written in block form as
\begin{equation}
  \begin{bmatrix}
    \bar{H}_{\mathrm{1h-1h}}   &\bar{H}_{\mathrm{1h-2h1p}}  \\
    \bar{H}_{\mathrm{2h1p-1h}} &\bar{H}_{\mathrm{2h1p-2h1p}}\\
  \end{bmatrix}
  \begin{bmatrix}
    C_{\mathrm{1h}}    \\
    C_{\mathrm{2h1p}}  \\
  \end{bmatrix} = E^{\mathrm{IP}}
  \begin{bmatrix}
    C_{\mathrm{1h}}    \\
    C_{\mathrm{2h1p}}  \\
  \end{bmatrix}.
  \label{IP_seq_eq}
\end{equation}
Here, $\bar{H}_{\mathrm{1h-1h}}$ is the ``one-hole -- one-hole'' block
involving $\bar{H}_{ij}$; $\bar{H}_{\mathrm{1h-2h1p}}$ and
$\bar{H}_{\mathrm{2h1p-1h}}$ are the ``one-hole -- two-hole-one-particle''
and ``two-hole-one-particle -- one-hole'' blocks involving $\bar{H}_{ij,ka}$
and $\bar{H}_{ka,ij}$, respectively; and $\bar{H}_{\mathrm{2h1p-2h1p}}$ is the
``two-hole-one-particle -- two-hole-one-particle'' block involving $\bar{H}_{ij}$,
$\bar{H}_{ab}$, $\bar{H}_{ij,kl}$, and $\bar{H}_{ia, bj}$. Similarly, the
eigenvalue equation for electron-attached states takes the block form
\begin{equation}
  \begin{bmatrix}
    \bar{H}_{\mathrm{1p-1p}}   &\bar{H}_{\mathrm{1p-1h2p}}  \\
    \bar{H}_{\mathrm{1h2p-1p}} &\bar{H}_{\mathrm{1h2p-1h2p}}\\
  \end{bmatrix}
  \begin{bmatrix}
    C_{\mathrm{1p}}    \\
    C_{\mathrm{1h2p}}  \\
  \end{bmatrix} = E^{\mathrm{EA}}
  \begin{bmatrix}
    C_{\mathrm{1p}}    \\
    C_{\mathrm{1h2p}}  \\
  \end{bmatrix}.
  \label{EA_seq_eq}
\end{equation}
Here, $\bar{H}_{\mathrm{1p-1p}}$ refers to the ``one-particle -- one-particle''
block involving $\bar{H}_{ab}$; $\bar{H}_{\mathrm{1p-1h2p}}$ and
$\bar{H}_{\mathrm{1h2p-1p}}$ are the ``one-particle - one-hole-two-particle''
and ``one-hole-two-particle -- one-particle'' blocks involving
$\bar{H}_{ci,ab}$ and $\bar{H}_{ab,ci}$, respectively; and
$\bar{H}_{\mathrm{1h2p-1h2p}}$ is the ``one-hole-two- particle --
one-hole-two-particle'' block involving
$\bar{H}_{ij}$, $\bar{H}_{ab}$, $\bar{H}_{ab,cd}$, and $\bar{H}_{ia, bj}$.

In contrast to CC theory, where $\bar{H}$ is non-Hermitian and its
commutator expansion terminates at {quartic power}, the $\bar{H}$ in
UCC theory is Hermitian and features a non-terminating commutator expansion.
We adopt a commutator expansion for $\bar{H}$ with Bernoulli-number prefactors,
\begin{align}
  \bar{H} &= \bar{H}^0 + \bar{H}^1 + \bar{H}^2 + \bar{H}^3 + \cdots, \label{hbar_total}\\
  \bar{H}^0 &= \hat{F} + \hat{V}, \label{hbar0} \\
  \bar{H}^1 &= [\hat{F}, \hat{\sigma}] + \frac{1}{2}[\hat{V}, \hat{\sigma}] +
  \frac{1}{2}[V_R, \sigma], \label{hbar1} \\
  \bar{H}^2 &= \frac{1}{12}[[\hat{V}_N, \hat{\sigma}],\hat{\sigma}] +
  \frac{1}{4}[[\hat{V}, \hat{\sigma}]_R, \hat{\sigma}] + \frac{1}{4} [ [
  \hat{V}_R, \hat{\sigma}]_R, \hat{\sigma}] \label{hbar2} \\
  \bar{H}^3 &= \frac{1}{24}[[[\hat{V}_N, \hat{\sigma}],\hat{\sigma}]_R, \hat{\sigma}] +
  \frac{1}{8}[[[\hat{V}_R, \hat{\sigma}]_R, \hat{\sigma}]_R, \hat{\sigma}] +
  \frac{1}{8}[[[\hat{V}, \hat{\sigma}]_R, \hat{\sigma}]_R, \hat{\sigma}] -
  \frac{1}{24}[[[\hat{V}, \hat{\sigma}]_R, \hat{\sigma}], \hat{\sigma}] - \nonumber \\
  &\phantom{=}\ \frac{1}{24}[[[\hat{V}_R, \hat{\sigma}]_R, \hat{\sigma}], \hat{\sigma}],
  \label{hbar3}
\end{align}
where ``$N$'' denotes the part of an operator consisting of excitation and
de-excitation components of the target operator, while ``$R$'' refers to the
reminder that excludes the ``$N$'' part. The Hamiltonian operator $\hat{H}$ is
decomposed into the Fock operator $\hat{F}$ and the fluctuation operator
$\hat{V}$, written in normal-ordered second-quantized form as
\begin{equation}
  \hat{F} = \sum_{ij} f_{ij} \{ \hat{a}_i^\dagger \hat{a}_j \} + \sum_{ab}
  f_{ab} \{ a_a^\dagger a_b \},  \quad
  \hat{V} = \frac{1}{4}\sum_{pqrs}\langle pq || rs\rangle \{
  \hat{a}_p^{\dagger}\hat{a}_q^\dagger \hat{a}_s \hat{a}_r \}.
\end{equation}
This Bernoulli expansion avoids all commutators of order higher than the single
commutator with respect to $\hat{F}$ and provides a compact form suitable for
practical UCCSD implementations. By applying L\"owdin
partitioning\cite{Lowdin1970} to eq.~\eqref{IP_seq_eq} and \eqref{EA_seq_eq},
the terms associated with the 2h1p (or 1h2p) excitations can be folded into
the 1h--1h (or 1p--1p) block. This yields to the following
energy-dependent eigenvalue equation for the 1h-1h (or 1p-1p) block:
\begin{eqnarray}
	\left[\bar{H}_{\mathrm{1h-1h}} + \bar{H}_{\mathrm{1h-2h1p}}
	\left(E^{\mathrm{IP}} - \bar{H}_{\mathrm{2h1p-2h1p}}\right)^{-1}\bar{H}_{\mathrm{2h1p-1h}} \right]
	C_{\mathrm{1h-1h}} &=& E^{\mathrm{IP}}C_{\mathrm{1h-1h}}, \\
	\left[\bar{H}_{\mathrm{1p-1p}} + \bar{H}_{\mathrm{1p-1h2p}}
	\left(E^{\mathrm{EA}} - \bar{H}_{\mathrm{1h2p-1h2p}}\right)^{-1}\bar{H}_{\mathrm{1h2p-1p}} \right]
	C_{\mathrm{1p-1p}} &=& E^{\mathrm{EA}}C_{\mathrm{1p-1p}}.
\end{eqnarray}
In these equations, $\bar{H}_{\mathrm{1h-1h}}$(or $\bar{H}_{\mathrm{1p-1p}}$)
primarily governs states dominated by the single-hole (or single-particle)
character. The term $\bar{H}_{\mathrm{1h-2h1p}}(E^{\mathrm{IP}} -
\bar{H}_{\mathrm{2h1p-2h1p}})^{-1}
\bar{H}_{\mathrm{2h1p-1h}}$ (or $\bar{H}_{\mathrm{1p-1h2p}}(E^{\mathrm{EA}} -
\bar{H}_{\mathrm{1h2p-1h2p}})^{-1} \bar{H}_{\mathrm{1h2p-1p}}$)
can be viewed as corrections to the leading block.
A balanced truncation scheme for practical UCC-EPT requires a {consistent
level of approximation} for both the dominant and correction terms. 
Specifically, the expansion terms retained in $\bar{H}_{\mathrm{1h-1h}}$
(or $\bar{H}_{\mathrm{1p-1p}}$) should be one order higher than
those in $\bar{H}_{\mathrm{1h-2h1p}}/\bar{H}_{\mathrm{2h1p-1h}}$
(or $\bar{H}_{\mathrm{1p-1h2p}}/\bar{H}_{\mathrm{1h2p-1p}}$), and two orders
higher than those in $\bar{H}_{\mathrm{2h1p-2h1p}}$ (or
$\bar{H}_{\mathrm{1h2p-1h2p}}$). To achieve third order accuracy within M\o
ller Plesset (MP) perturbation theory for single-hole
(or single-particle)-dominated states, the terms in $\bar{H}_{\mathrm{1h-1h}}$ (or
$\bar{H}_{\mathrm{1p-1p}}$) should be evaluated through third order, the
terms in $\bar{H}_{\mathrm{1h-2h1p}}/\bar{H}_{\mathrm{2h1p-1h}}$
(or $\bar{H}_{\mathrm{1p-1h2p}}/\bar{H}_{\mathrm{1h2p-1p}}$) through second
order, and $\bar{H}_{\mathrm{2h1p-2h1p}}$ (or $\bar{H}_{\mathrm{1h2p-1h2p}}$)
through first order. In addition, the two UCCSD amplitude equations,
$\bar{H}_{ai}=0$ and $\bar{H}_{ab,ij}=0$, used to iteratively optimize
the ground-state amplitudes, can also be evaluated through third order.
By analogy with UCC3 in UCC-PPT\cite{Liu2018}, this formulation is termed
the IP/EA-UCC3 scheme. For further simplification, substituting the MP3
ground-state amplitudes directly into IP/EA-UCC3 yields a perturbative
variant -- the strict version of IP/EA-UCC3 (IP/EA-UCC3-s), which is
equivalent to the IP/EA-ADC(3) method.\cite{Hodecker2022,Hodecker2025}
An alternative and natural truncation strategy for UCC-based methods is
truncation by commutator order. One of the simplest such models in UCC-PPT is
the quadratic UCCSD (qUCCSD) scheme.\cite{Liu2021} Following this strategy,
the IP/EA-qUCCSD scheme is readily formulated for calculating IPs and EAs.
In IP/EA-qUCCSD, both the ground-state amplitude equations and
$\bar{H}_{\mathrm{1h-1h}}$(or $\bar{H}_{\mathrm{1p-1p}}$) block in the
electron-detached(or -attached) state eigenvalue equations are truncated to
include double commutators between $\hat{V}$ and $\hat{\sigma}$ in the
Bernoulli expansion. Simultaneously, $\bar{H}_{\mathrm{1h-2h1p}}$ and
$\bar{H}_{\mathrm{2h1p-1h}}$ (or $\bar{H}_{\mathrm{1p-1h2p}}$ and
$\bar{H}_{\mathrm{1h2p-1p}}$) are truncated to the single-commutator level,
while $\bar{H}_{\mathrm{2h1p-2h1p}}$ (or $\bar{H}_{\mathrm{1h2p-1h2p}}$)
is truncated to the bare Hamiltonian. Compared with IP/EA-UCC3 scheme, the
IP/EA-qUCCSD approach includes additional higher-order contributions arising
from the commutator structure in both the amplitude equations and
electron-detached (or attached)-state eigenvalue equations. In principle, this
leads to a systematic improvement over the IP/EA-UCC3 method.

\subsection{The working equations for the IP/EA-qUCCSD scheme}
The working equations for the IP/EA-qUCCSD method are derived using
the Bernoulli expansion of $\bar{H}$ (as outlined in the previous subsection),
together with diagrammatic techniques analogous to those employed
in UCC-PPT. The expressions for the ground-state energy and the two
amplitude equations are identical to those in qUCCSD.\cite{Liu2021} For completeness,
the relevant formulas are collected in the Appendix. Here we focus on the contributions
to $\bar{H}$ that enter the various blocks of the electron-detachment and
electron-attachment [($N\pm1$)-electron state] eigenvalue equations.

In the IP-qUCCSD approach, the ``1h-1h'' block of the electron-detachment eigenvalue
equation arises solely from $\bar{H}_{ij}$, truncated at the double-commutator level.
The resulting expressions for $\bar{H}_{ij}^{\mathrm{IP-qUCCSD}}$ can be decomposed as
\begin{eqnarray}
  \bar{H}_{ij}^{\mathrm{IP-qUCCSD}} &=& \bar{H}_{ij}^{0} + \bar{H}_{ij}^{1} + \bar{H}_{ij}^{2},\\
        \bar{H}_{ij}^{0} &=& f_{ij}, \\
        \bar{H}_{ij}^{1} &=&\sum_{kab}\frac{1}{4}\langle
    ik\|ab\rangle\sigma_{jk}^{ab}+\sum_{ka}\langle
    ik\|ja\rangle\sigma_{k}^{a}+h.c., \\
        \bar{H}_{ij}^{2} &=&
    \left(\sum_{klabc}\frac{1}{2}(\sigma_{kl}^{bc})^{*}\langle
    ic\|al\rangle\sigma_{jk}^{ab}+\sum_{klmab}\frac{1}{8}(\sigma_{kl}^{ab})^{*}\langle
    im\|kl\rangle\sigma_{jm}^{ab}+h.c.\right) \nonumber \\
    &-&\sum_{klmab}\frac{1}{2}(\sigma_{kl}^{ab})^{*}\langle
    im\|jl\rangle\sigma_{km}^{ab}+\sum_{klabc}\frac{1}{2}(\sigma_{kl}^{ac})^{*}\langle
    ic\|jb\rangle\sigma_{kl}^{ab}  \nonumber \\
        &+&\left(\sum_{kabc}\frac{1}{4}(\sigma_{k}^{b})^{*}\langle
    ib\|ac\rangle\sigma_{jk}^{ac}-\sum_{klab}\frac{1}{2}(\sigma_{k}^{b})^{*}\langle
    il\|ak\rangle\sigma_{jl}^{ab}+\sum_{klab}\frac{1}{2}(\sigma_{l}^{b})^{*}\langle
    ik\|ja\rangle\sigma_{kl}^{ab}+h.c.\right)  \nonumber \\
        &+&\left(\sum_{kab}\frac{5}{12}\langle
    ik\|ab\rangle\sigma_{j}^{a}\sigma_{k}^{b}+\sum_{kab}\frac{1}{2}(\sigma_{k}^{b})^{*}\langle
    ib\|ak\rangle\sigma_{j}^{a}+h.c.\right) \nonumber \\
        &-&\sum_{kla}(\sigma_{l}^{a})^{*}\langle
    ik\|jl\rangle\sigma_{k}^{a}+\sum_{kab}(\sigma_{k}^{a})^{*}\langle
    ia\|jb\rangle\sigma_{k}^{b}.
    \end{eqnarray}
The $\bar{H}_{\mathrm{1h-2h1p}}$ block contains only $\bar{H}^{\mathrm{IP-qUCCSD}}_{ij,ka}$,
expanded to the single-commutator level:
   \begin{eqnarray}
    \bar{H}_{ij,ka}^{\mathrm{IP-qUCCSD}} &=& \bar{H}_{ij,ka}^{0} + \bar{H}_{ij,ka}^{1}, \\
    \bar{H}_{ij,ka}^{0} &=& \langle ij\|ka\rangle,  \\
    \bar{H}_{ij,ka}^{1} &=&  P(ij)\sum_{lb}(\sigma_{jl}^{ab})^{*}\langle ib\|kl\rangle
      + \sum_{bc}\frac{1}{2}(\sigma_{ij}^{cb})^{*}\langle bc\|ak\rangle 
      + \sum_{b}\frac{1}{2}\langle ij\|ba\rangle\sigma_{k}^{b}  \nonumber \\
     &-& \sum_{l}(\sigma_{l}^{a})^{*}\langle ij\|kl\rangle -
       P(ij)\sum_{b}(\sigma_{j}^{b})^{*}\langle ib\|ak \rangle.
   \end{eqnarray}
The $\bar{H}_{\mathrm{2h1p-1h}}$ block contains $\bar{H}_{ka,ij}$, the Hermitian conjugate
of $\bar{H}_{ij,ka}$, i.e., $\bar{H}_{ij,ka} = (\bar{H}_{ka,ij})^*$.
In addition to $\bar{H}^0_{ij}$ disscussed above, the $\bar{H}_{\mathrm{2h1p-2h1p}}$ block also contains
the bare Hamiltonian terms corresponding to $\bar{H}_{ab}^0$, $\bar{H}_{ia,bj}^0$,
and $\bar{H}_{ij,kl}^0$, which are given by
\begin{eqnarray}
  \bar{H}^{\mathrm{IP-qUCCSD[0]}}_{ab} &=& \bar{H}_{ab}^0 = f_{ab}, \\
	\bar{H}^{\mathrm{IP-qUCCSD}}_{ia, bj} &=& \bar{H}_{ia, bj}^0 = \langle ia \| bj \rangle, \\
	\bar{H}^{\mathrm{IP-qUCCSD}}_{ij, kl} &=& \bar{H}_{ij, kl}^0 = \langle ij \| kl \rangle. 
\end{eqnarray}

\begin{figure}[!hbtp]
  \captionsetup[subfigure]{labelformat=empty}
  \centering
  \begin{subfigure}[b]{0.4\textwidth}{
    \centering
    \includegraphics[width=\textwidth]{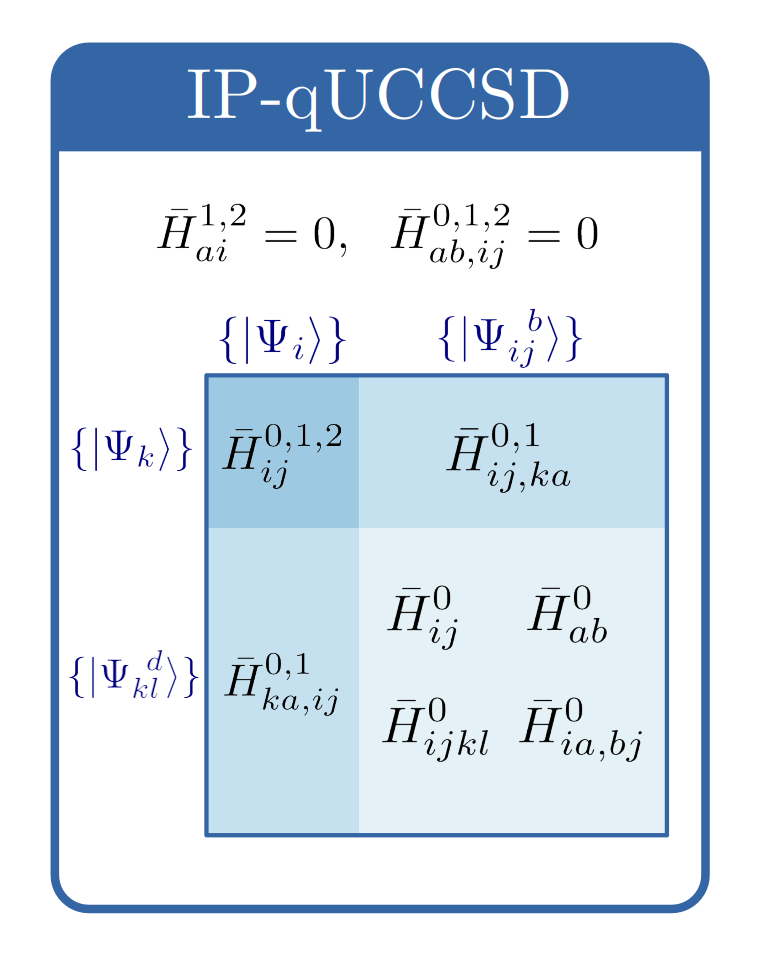}
    \label{IP-qUCCSD}}
  \end{subfigure} \hspace{3em}
  \begin{subfigure}[b]{0.4\textwidth}{
    \centering
    \includegraphics[width=\textwidth]{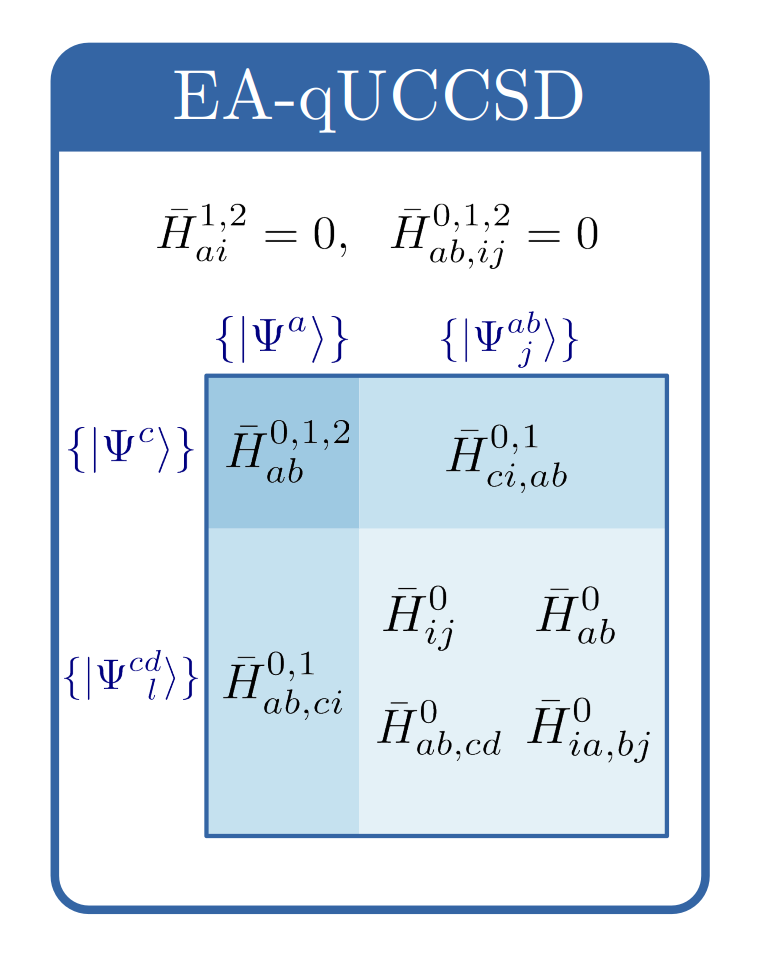}
    \label{EA-qUCCSD}}
  \end{subfigure}
  \begin{subfigure}[b]{0.4\textwidth}{
    \centering
    \includegraphics[width=\textwidth]{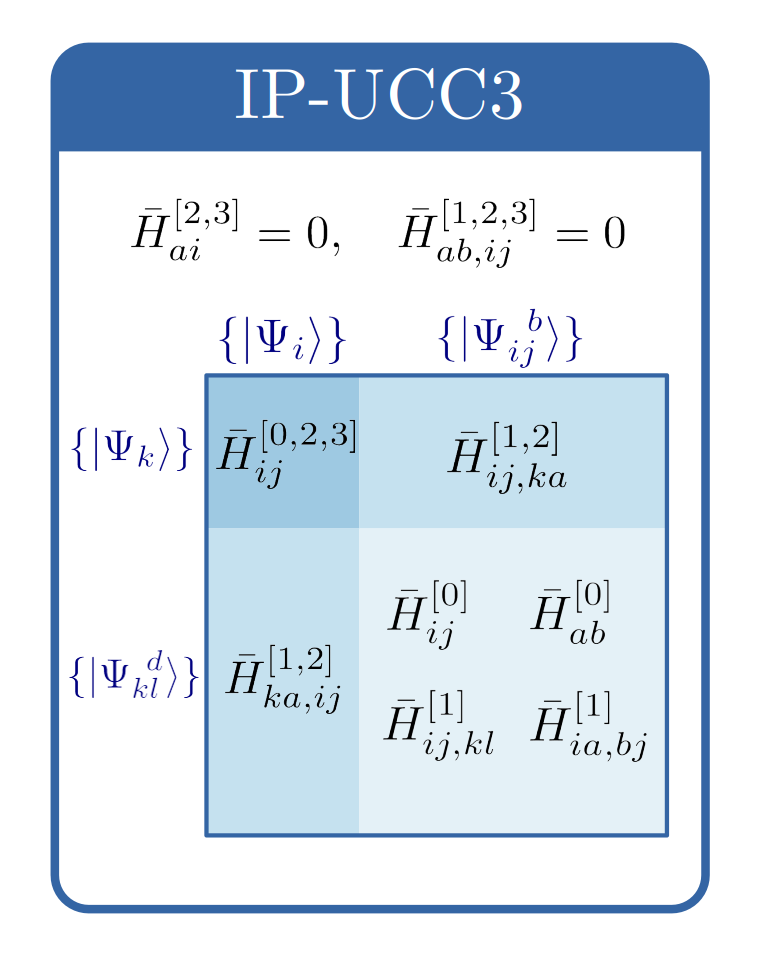}
    \label{IP-UCC3}}
  \end{subfigure} \hspace{3em}
  \begin{subfigure}[b]{0.4\textwidth}{
    \centering
    \includegraphics[width=\textwidth]{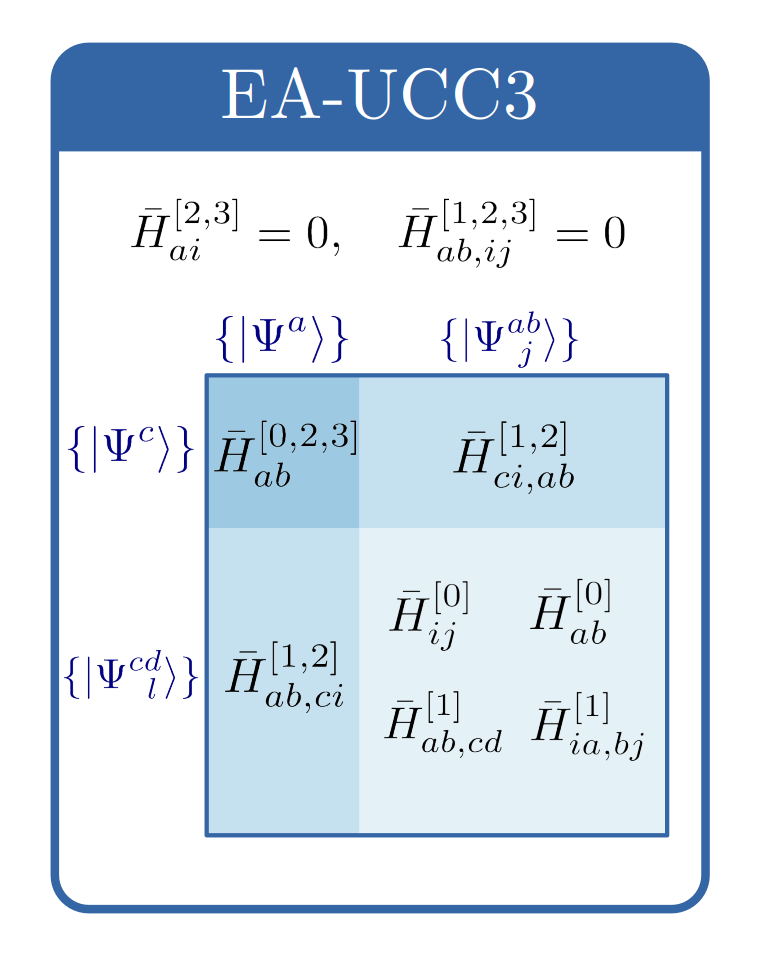}
    \label{EA-UCC3}}
  \end{subfigure}
  \vspace{-1em}
  \caption{Block structure and constituent terms of the similarity-transformed
  Hamiltonian ($\bar{H}$) and amplitude equations in practical UCC-based
  electron propagator implementations. In the UCC3 scheme, the superscripts on
  $\bar{H}$ denote the perturbation order to which both $\bar{H}$ and the
  amplitude equations are expanded in each block. In the qUCCSD scheme, the
  superscripts instead indicate commutator rank, as defined in eq~\eqref{hbar_total}
  -\eqref{hbar3}.}
  \label{Schematic_pic}
\end{figure}

Similarly, in the EA-qUCCSD scheme, the ``1p-1p'' block of the electron-attachment
eigenvalue equation contains only contributions from $\bar{H}_{ab}$, truncated
at the double-commutator level. The corresponding expressions are
\begin{align}
  \bar{H}_{ab}^{\mathrm{EA-qUCCSD}} &= \bar{H}_{ab}^{0} + \bar{H}_{ab}^{1} +
  \bar{H}_{ab}^{2},\\
    \bar{H}_{ab}^{0} &= f_{ab}, \\
  \bar{H}_{ab}^{1} &= -\sum_{ijc} \frac{1}{4}\langle i j \| b c\rangle
  \sigma_{ij}^{ac}+\sum_{ic}\langle a i \| b c\rangle \sigma_{i}^{c}+h.c., \\ 
  \bar{H}_{ab}^{2} &= \left(-\sum_{ijkcd} \frac{1}{2}
  \left(\sigma_{i j}^{c d}\right)^{*}\langle kd \| bj\rangle \sigma_{i k}^{c a}
  -\sum_{ijcdf} \frac{1}{8}\left(\sigma_{ij}^{fd}\right)^{*}\langle d f \| c b\rangle 
  \sigma_{i j}^{a c}+h. c.\right) \nonumber \\
    &+\sum_{i j c d f} \frac{1}{2}\left(\sigma_{i j}^{f
  d}\right)^{*}\langle a d \| b c\rangle \sigma_{i j}^{f c}-\frac{1}{2}
  \sum_{i j k c d}\left(\sigma_{i j}^{c d}\right)^{*}\langle k a \| j
  b\rangle \sigma_{i k}^{c d} \nonumber \\
  &+\left(\sum_{i j k c}
  \frac{1}{4}\left(\sigma_{j}^{c}\right)^{*}\langle i k \| b j\rangle
  \sigma_{i k}^{a c}-\sum_{i j c d}
  \frac{1}{2}\left(\sigma_{j}^{c}\right)^{*}\langle i c \| b d\rangle
  \sigma_{i j}^{a d}+\sum_{i j c d}
  \frac{1}{2}\left(\sigma_{j}^{d}\right)^{*}\langle i a \| c b\rangle
  \sigma_{i j}^{c d}+h. c.\right) \nonumber\\ 
  &+\left(-\frac{5}{12} \sum_{i j
  c}\langle i j \| b c\rangle \sigma_{i}^{a} \sigma_{j}^{c}-\sum_{i j c}
  \frac{1}{2}\left(\sigma_{j}^{c}\right)^{*}\langle i c \| b j\rangle
  \sigma_{i}^{a}+h. c.\right) \nonumber \\
  &-\sum_{i j c}\left(\sigma_{i}^{c}\right)^{*}\langle j a \| i b\rangle
  \sigma_{j}^{c}+\sum_{i c d}\left(\sigma_{i}^{d}\right)^{*}\langle a d \|
  b c\rangle \sigma_{i}^{c}.
\end{align}
The $\bar{H}_{\mathrm{1p-1h2p}}$ block includes only $\bar{H}^{\mathrm{EA-qUCCSD}}_{ci,ab}$,
expanded up to the single-commutator level. It is the Hermitian conjugate of
$\bar{H}_{ab,ci}^{\mathrm{EA-qUCCSD}}$, which contributes to the $\bar{H}_{\mathrm{1h2p-1p}}$ block: 
\begin{eqnarray}
  \bar{H}_{ab,ci}^{\mathrm{EA-qUCCSD}} &=& \bar{H}_{ab,ci}^{0} + \bar{H}_{ab,ci}^{1}, \\
    \bar{H}_{ab,ci}^{0} &=& \langle ab\|ci\rangle, \\
    \bar{H}_{ab,ci}^{1} &=& P(ab)\sum_{jd}\langle aj\|cd\rangle\sigma_{ij}^{bd}
    + \sum_{kj}\frac{1}{2}\langle jk\|ci\rangle\sigma_{jk}^{ab} -
    \sum_{j}\frac{1}{2}(\sigma_{j}^{c})^{*}\langle ab\|ji\rangle \nonumber \\
    &+&\sum_{d}\langle ab\|cd\rangle\sigma_{i}^{d} - P(ab)\sum_{j}\langle
  aj\|ci\rangle\sigma_{j}^{b}.
\end{eqnarray}
The $\bar{H}_{\mathrm{1h2p-1h2p}}$ consists of the bare Hamiltonian contributions
$\bar{H}_{ab}^0$, $\bar{H}_{ij}^0$, $\bar{H}_{ab,cd}^0$, and $\bar{H}_{ia,bj}^0$. 
In fact, $\bar{H}_{ab}^0$, $\bar{H}_{ij}^0$, and $\bar{H}_{ia,bj}^0$ in
$\bar{H}_{\mathrm{1h2p-1h2p}}$ block are identical to those in
$\bar{H}_{\mathrm{2h1p-2h1p}}$ block:
\begin{equation}
  \bar{H}_{ab}^{\mathrm{IP-qUCCSD}} = \bar{H}_{ab}^{\mathrm{EA-qUCCSD[0]}} = f_{ab},\
  \bar{H}_{ij}^{\mathrm{IP-qUCCSD[0]}} = \bar{H}_{ij}^{\mathrm{EA-qUCCSD}},\
  \bar{H}_{ia,bj}^{\mathrm{IP-qUCCSD}} = \bar{H}_{ia,bj}^{\mathrm{EA-qUCCSD}}.
\end{equation}
The $\bar{H}_{ab,cd}^{\mathrm{EA-qUCCSD}}$ contributes exclusively to
$\bar{H}_{\mathrm{2h1p-2h1p}}$ and equals the bare Hamiltonian,
\begin{equation}
  \bar{H}_{ab,cd}^{\mathrm{EA-qUCCSD}} = \bar{H}_{ab,cd}^0 = \langle ab \| cd \rangle.
\end{equation}
It is worth noting that the amplitude equations and target-state
eigenvalue equations in the IP/EA-UCC3 scheme constitute a subset of those in
IP/EA-qUCCSD. Thus, IP/EA-UCC3 can be straightforwardly obtained by omitting
terms beyond third order in the qUCCSD ground-state amplitude equations and
in the $\bar{H}^{\mathrm{IP-qUCCSD}}_{\mathrm{1h-1h}}$/
$\bar{H}^{\mathrm{EA-qUCCSD}}_{\mathrm{1p-1p}}$ block;
terms beyond second order in the
$\bar{H}^{\mathrm{IP-qUCCSD}}_{\mathrm{1h-2h1p}}(\bar{H}^{\mathrm{IP-qUCCSD}}_{\mathrm{2h1p-1h}})$/
$\bar{H}^{\mathrm{EA-qUCCSD}}_{\mathrm{1p-1h2p}}(\bar{H}^{\mathrm{EA-qUCCSD}}_{\mathrm{1h2p-1p}})$
blocks; and terms beyond first order in the $\bar{H}^{\mathrm{IP-qUCCSD}}_{\mathrm{2h1p-2h1p}}$
/$\bar{H}^{\mathrm{EA-qUCCSD}}_{\mathrm{1h2p-1h2p}}$ block.
The connection between IP/EA-UCC3 and IP/EA-ADC(3) has been discussed in
refs.~\citenum{Hodecker2022} and \citenum{Hodecker2025}.
Figure~\ref{Schematic_pic} illustrates the block structure and constituents of
$\bar{H}$ in the electron-attachment and electron-detachment eigenvalue equations
discussed above. For comparison, the corresponding IP/EA-UCC3 block structure
and components are also shown.

\subsection{Relation between working equations of UCC-EPT and UCC-PPT}
At the level of working equations, UCC-based EPT and PPT share the same ground-state
amplitude equations for a given truncation scheme. Moreover, the
structure of the ($N\pm1$)-electron state eigenvalue equations in UCC-EPT closely
parallels that of the excited-state eigenvalue equations in UCC-PPT.
Within the UCCSD model, both frameworks yield $2 \times 2$ block-type eigenvalue
equations. {In the qUCCSD scheme,\cite{Liu2021} the $\bar{H}_{\mathrm{SS}}$ block 
provides the largest contribution to excitation energies dominated by single-excitation
character; likewise, the $\bar{H}_{\mathrm{1h-1h}}$ and $\bar{H}_{\mathrm{1p-1p}}$ 
blocks contribute most to the IPs of 1h-dominated electron-detached states and to the EAs
of 1p-dominated electron-attached state, respectively.}
Hence, each corresponding block in the $2\times 2$ block matrices serves a
comparable role within the eigenvalue equations of the two frameworks. 
The key distinction between UCC-EPT and UCC-PPT lies in the specific terms
included in the respective blocks: each block in the ($N\pm1$)-electron state
eigenvalue equations of UCC-EPT contains a subset of the terms present in the
corresponding excited-state eigenvalue equations of UCC-PPT.  The relevant
$\bar{H}$ components of the qUCCSD and IP/EA-qUCCSD eigenvalue equations are
summarized in Table~\ref{t1}. In the IP-qUCCSD scheme, the $\bar{H}_{\mathrm{1h-1h}}$,
$\bar{H}_{\mathrm{1h-2h1p}}$, and $\bar{H}_{\mathrm{2h1p-1h}}$ blocks draw
their contributions solely from $\bar{H}_{ij}$, $\bar{H}_{ij,ka}$, and
$\bar{H}_{ci,ab}$, respectively -- terms
that also constitute parts of $\bar{H}_{\mathrm{SS}}$, $\bar{H}_{\mathrm{SD}}$ and
$\bar{H}_{\mathrm{DS}}$ in qUCCSD. The $\bar{H}_{\mathrm{DD}}$ block
in qUCCSD includes an additional $\bar{H}_{ab,cd}$ contribution that is absent
from the $\bar{H}_{\mathrm{2h1p-2h1p}}$ block of IP-qUCCSD.
Analogous relationships hold for the EA-qUCCSD method.
Consequently, once UCC-PPT equations are derived and implemented, the
corresponding UCC-EPT formulations can be constructed directly and
efficiently, since all required $\bar{H}$ terms are already available within the
UCC-PPT framework.

The overall formal scalings of the ground-state amplitude equations and the
($N\pm1$)-electron state eigenvalue equations in the IP/EA-qUCCSD method are
$\mathcal{O}(N^6)$ and $\mathcal{O}(N^5)$, respectively -- identical to those of
IP/EA-EOM-CCSD. Because the ground-state theory of IP/EA-qUCCSD is
the same as that of qUCCSD, the computational cost of solving the ground-state
amplitude equations is approximately twice that of a CCSD calculation\cite{Liu2022}.
Compared with IP/EA-ADC(3) (also referred to as IP/EA-UCC3-s), both IP/EA-qUCCSD
and IP/EA-UCC3 are substantially more demanding, as they require iterative
optimizations of the ground-state amplitudes rather than using perturbative
amplitudes. The solutions of the eigenvalue equations for IP- and EA-qUCCSD scale
as $\mathcal{O}(N_o^3N_v^2)$ and $\mathcal{O}(N_oN_v^4)$, respectively, matching
the formal scaling of IP/EA-EOM-CCSD.
{In practice, IP/EA-qUCCSD is slightly less computationally expensive than
IP/EA-EOM-CCSD. In IP/EA-EOM-CCSD, the similarity-transformed Hamiltonian $\bar{H}$
contains three-body terms -- namely,\\
$\frac{1}{2}\sum_{mnef}t_{ij}^{ae} \langle mn || ef \rangle r_{mn}^f $
in the $\bar{H}_{\mathrm{2h1p-2h1p}}$ block and
$-\frac{1}{2}\sum_{klcd} t_{kj}^{ab} \langle kl || cd \rangle r_{\phantom{i}l}^{cd}$
in the $\bar{H}_{\mathrm{1h2p-1h2p}}$ block -- both arising from $[\hat{V},\hat{T}_2]$
in the BCH expansion. By contrast, in IP/EA-qUCCSD the $\bar{H}_{\mathrm{2h1p-2h1p}}$
and $\bar{H}_{\mathrm{1h2p-1h2p}}$ blocks are constructed directly from the
bare Hamiltonian in the Bernoulli expansion of the $\bar{H}^{\mathrm{IP/EA-qUCCSD}}$; 
thus, no three-body terms appear in its eigenvalue equations. Moreover,
coupling terms such as $\sum_{me} \bar{H}_{me} r_{mi}^e$ in $\bar{H}_{\mathrm{1h-2h1p}}$ block and 
$\sum_{me}\bar{H}_{me}r_{\phantom{i}m}^{ae}$ in $\bar{H}_{\mathrm{1p-1h2p}}$ block --
present in IP/EA-EOM-CCSD -- do not contribute in the UCC-based PPT/EPT formulation as well. 
This follows from the Hermicity of $\bar{H}^{\mathrm{UCC}}$ together with the amplitude equations, 
which enforce the vanishing of the $\bar{H}_{me}^{\mathrm{qUCCSD}}$. Consequently,
IP/EA-qUCCSD avoids these contributions, leading to a lower overall computational cost
compared to IP/EA-EOM-CCSD.}
The construction of the similarity-transformed Hamiltonian $\bar{H}$ in the IP/EA-qUCCSD also scales as
$\mathcal{O}(N^6)$, as in qUCCSD. In contrast, UCC-EPT calculations are
generally less computationally intensive than qUCCSD, since fewer $\bar{H}$
terms appear in the various blocks of the eigenvalue equations --
particularly in the {cost-dominating} blocks $\bar{H}_{\mathrm{1h-1h}}$ for
IP-qUCCSD and $\bar{H}_{\mathrm{1p-1p}}$ for EA-qUCCSD. On the other hand,
IP/EA-qUCCSD remains somewhat more expensive than IP/EA-EOM-CCSD because
additional $\mathcal{O}(N_o^4 N_v^2)$ and $\mathcal{O}(N_o^3 N_v^3)$ terms are
required when constructing $\bar{H}$, especially in these {cost-dominating} blocks.
Nevertheless, these $\mathcal{O}(N^6)$-scaling intermediates are computed only
once and can be reused in subsequent solutions of the eigenvalue equations.

\section{Computational details}
The IP/EA-UCC3 and IP/EA-qUCCSD methods were implemented in the PySCF software
package\cite{Sun2020_pyscf} using generalized orbital formulation.
Building on existing UCC3 and qUCCSD implementations, the programs for
IP/EA-UCC3 and IP/EA-qUCCSD described in section 2 were reorganized
and extended. The correctness of the UCC3 and qUCCSD implementations in PySCF
was verified by comparison with our earlier implementation in the X2CSOCC
module\cite{Liu2018_X2CSOCC} of the CFOUR program.\cite{cfour2020,cfour}
Benchmark calculations of one-hole (1h)-dominated {vertical IPs (VIPs)} were performed using the
dataset from ref.~\citenum{Dempwolff2020-IP-II} and {compared with}
IP-EOM-CCSD, IP-ADC(3), IP-ADC(4), and IP-UCC3. This benchmark set comprises 25
{VIPs} from closed-shell systems (six neutral molecules and four anions)
and 17 {VIPs} from open-shell systems (two radicals, three neutral triplet molecules, 
and four radical anions), providing a broad and representative
assessment. Reference {VIPs} were obtained from full configuration interaction
(FCI) calculations, with the exception of \ce{CO}, \ce{HCN}, \ce{NO2-},
\ce{NO2^{$\bullet$}} and \ce{HCN^{$\bullet$-}}, which were calculated using 
configuration interaction with singles, doubles, triples, and quadruples (CISDTQ). 
In addition, a dataset\cite{Ranasinghe2019} of 201 1h-dominated {VIPs} was used
for further benchmarking. In this dataset, high-level reference values were
computed using the IP-EOM-CCSDT method. The set spans 42 molecules with
closed-shell singlet ground states, offering a robust statistical basis for
error analysis. Benchmark calculations of {vertical EAs (VEAs)} dominated by one-particle
(1p)-dominated {VEAs} were conducted using the dataset from
ref.~\citenum{Dempwolff2021-EA} and compared with EA-EOM-CCSD, EA-ADC(3),
EA-ADC(4), and EA-UCC3. This {VEA} benchmark dataset includes 35 {VEAs}
from 10 closed-shell systems (seven neutral molecules and three cations)
containing up to three second-row atoms, as well as 16 {VEAs} from 8 open-shell
systems (two radicals, three neutral triplet molecules, and three
radical cations). Reference {VEAs} were obtained using FCI or, for \ce{CO},
\ce{HCN}, \ce{NO2+} and \ce{NO2^{$\bullet$}}, CISDTQ.
To enable comparison with existing ADC benchmarks, all datasets were
selected to match the statistical scopes reported in prior studies.
\cite{Dempwolff2020-IP-II,Dempwolff2021-EA,Leitner2024}
In all calculations, {standard non-relativistic two-electron integrals were used 
without approximation, and} the $1s$ orbitals of non-hydrogen atoms were kept frozen.
Detailed geometries of all test systems and the basis sets employed are
provided in the Supporting Information (SI). {They can also be found in the original
references.}\cite{Dempwolff2020-IP-II, Ranasinghe2019, Dempwolff2021-EA}.

\section{Results and discussion}
\subsection{Benchmark calculations of ionization potentials (IPs)}
\begin{figure}[!htbp]
  \centering
  \includegraphics[width=0.9\textwidth]{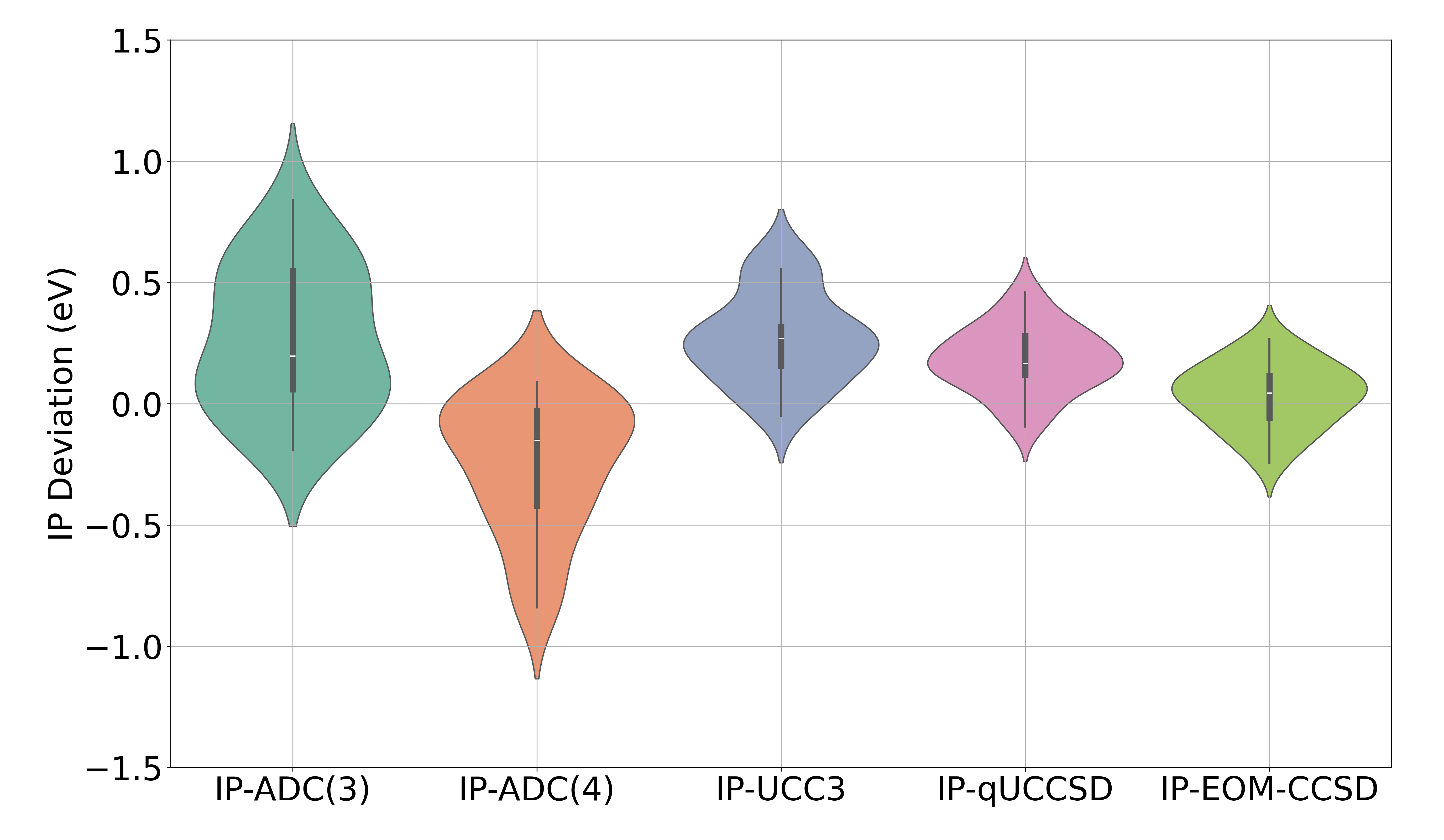}
  \caption{Violin plot of 25 1h-dominated {vertical} ionization potentials obtained for systems with closed-shell reference.}
  \label{IP_violin}
\end{figure}

A statistical analysis of the deviations in 25 IPs for closed-shell reference
states, calculated by IP-qUCCSD, IP-UCC3, IP-ADC(3), IP-ADC(4), and IP-EOM-CCSD
relative to FCI reference data, is presented in Table~\ref{t2}.
The corresponding vertical IPs computed with these methods are listed in
Table~S1 of the Supporting Information. For Hermitian excited-state methods,
the two UCC-based approaches generally outperform the ADC counterparts for
1h-dominated electron-detached states. This trend contrasts with that observed for
excitation energies (EEs), where qUCCSD performs {nearly} identically to ADC(3).
\cite{Liu2022,Liu2025} Specifically, IP-UCC3 yields a mean absolute deviation
({$\mathrm{MAD} = \frac{1}{n}\sum_i^n |\Delta_i|,\ \Delta_i = E_i^{\mathrm{calc}}-E_i^{\mathrm{FCI}}$})
comparable to IP-ADC(4) but with a lower standard deviation ({$\mathrm{SD}=
\sqrt{\frac{1}{n}\sum_i^n(\Delta_i - \bar{\Delta})^2}$}), whereas
IP-qUCCSD further improves upon IP-ADC(4), reducing the MAD and SD by 30\% and
52\%, respectively. In principle, IP-ADC(3), IP-UCC3 and IP-qUCCSD can all be
regarded as ``third-order methods'', capable of achieving IPs accurate to the
third order. In IP-qUCCSD, iterative optimization of ground-state amplitudes,
together with fourth- and fifth-order contributions entering via the double
commutator in the electron-detachment eigenvalue equation, yields a noticeably
improved description of 1h-dominated {VIPs} compared to IP-ADC(3).
From a theoretical perspective, one would therefore expect IP-ADC(4) -- a
formal fourth-order method -- to surpass both IP-UCC3 and IP-qUCCSD in accuracy.
However, IP-qUCCSD clearly outperforms IP-ADC(4) despite omitting
triple-excitation contributions and the fourth-order terms arising from the
triple commutators. A comparison of the mean deviation 
({$\mathrm{MD}=\frac{1}{n}\sum_i^n \Delta_i,\ \Delta_i = E^\mathrm{calc}_i - E^{\mathrm{FCI}}_i$}),
maximum positive deviation {$(\mathrm{MaxD}(+) = \mathrm{max}(\{\Delta_i\}))$},
and minimum negative deviation {$(\mathrm{MinD}(-)= \mathrm{min}(\{\Delta_i\})$)}
between IP-ADC(3) and IP-ADC(4), along with their
{median deviations and overall distributions depicted in the violin plots} 
(Figure~\ref{IP_violin}), shows that IP-ADC(3) tends to overestimate VIPs, whereas IP-ADC(4) tends
to underestimate them. When IP-ADC(2) and other IP-ADC(4) variants are also
considered,\cite{Leitner2024} an oscillatory convergence pattern emerges within
the ADC family, with ADC(4) apparently overcorrecting the deficiencies of ADC(3).
{This behavior suggests that simple inclusion of higher-order terms in the MP
perturbation expansion for the amplitudes does not necessarily yield a
systematic improvement in accuracy for the ADC family of methods. Instead of
a fixed-order MP expansion of the amplitudes, the commutator-truncation scheme
in qUCCSD treats the cluster amplitudes consistently and is more compatible with
the self-consistent solution of the amplitude equations. Collectively, these
considerations may explain the unexpectedly superior performance relative to
IP-ADC(4).}

Similar to the UCC3 versus qUCCSD case,\cite{Liu2021,Liu2025} IP-qUCCSD
consistently improves upon IP-UCC3, reducing the MAD and SD by 0.06 eV and
{0.05} eV, respectively. Among the four Hermitian methods evaluated,
IP-qUCCSD delivers the best overall numerical performance. This conclusion is
further supported by benchmark calculations of 201 {VIPs} with closed-shell
reference states from the CGB dataset reported in Ref~\citenum{Ranasinghe2019}.
The complete list of {VIPs} computed using the examined methods is provided in
Table~S4 of the Supporting Information. As shown in Table~\ref{t3}, the deviation
statistics increase for all methods on the CGB dataset. The expanded
dataset and the use of IP-EOM-CCSDT reference values significantly affect
the deviation distributions, particularly by amplifying both MaxD(+) and
MinD(-) across methods. Despite the resulting increases in MAD and SD for the
UCC-based approaches, IP-UCC3 performs comparably to IP-ADC(4), while IP-qUCCSD
maintains a clear advantage, with MAD and SD values lower than those of
IP-ADC(4) by 0.04 eV and 0.05 eV, respectively. Overall, IP-qUCCSD remains
the most accurate of the four Hermitian methods examined. On the 25-{VIP} dataset,
IP-EOM-CCSD attains slightly higher accuracy than IP-qUCCSD, with a lower MAD
of 0.11 eV and the same SD of 0.13 eV. When applied to the CGB
dataset, both IP-EOM-CCSD and IP-qUCCSD show increased deviations, with a more
pronounced deterioration for IP-EOM-CCSD (its MAD and SD more than double relative
to the 25-{VIP} set). In this case, IP-qUCCSD matches the MAD of IP-EOM-CCSD but
achieves a lower SD of 0.22 eV and exhibits less severe negative outlier.
Within the exception of IP-ADC(4), all methods tend to overestimate the
vertical IPs. The MD, MaxD(+) and MinD(-) values for IP-UCC3 and IP-qUCCSD
reveal a decreasing degree of overestimation, further underscoring the
systematic improvement from IP-UCC3 to IP-qUCCSD. Taken together, these
statistical results indicate that the performance of IP-qUCCSD approaches that
of IP-EOM-CCSD. This finding highlights the potential for further development
of more accurate methods within the UCC-EPT framework, both to explore the
limitations of the current UCCSD model and to clarify the correspondence
between the UCCSD truncation scheme and IP-EOM-CCSD accuracy.

By utilizing different reference states, IP/EA methods can target $(N\pm
1)$-electron states that differ by a single electron from the reference.
It is therefore essential to assess {the performance of UCC-EPT} for systems with
open-shell references. Table~\ref{t4} summarizes the statistical
deviations for 17 1h-character {VIPs} with open-shell reference from
ref.~\citenum{Dempwolff2020-IP-II}, computed by IP-qUCCSD, IP-UCC3, IP-ADC(3),
IP-ADC(4), and IP-EOM-CCSD relative to FCI. The full list of calculated {VIPs}
is provided in Table~S3 of the Supporting Information. Among the five methods,
IP-ADC(4) delivers the highest accuracy and precision, reflecting the important
role of triple excitations in computing {VIPs} for open-shell reference systems.
For IP-qUCCSD, the MAD and SD exceed those of IP-ADC(4) both by about 0.03 eV,
and its MaxD(+) is substantially larger. Nevertheless, a slight improvement
over IP-UCC3 persists for open-shell systems. Overall, IP-qUCCSD performs
nearly on par with IP-EOM-CCSD and clearly surpasses IP-ADC(3).

\subsection{Benchmark calculations of electron affinities (EAs)}
\begin{figure}[!htbp]
  \centering
  \includegraphics[width=0.9\textwidth]{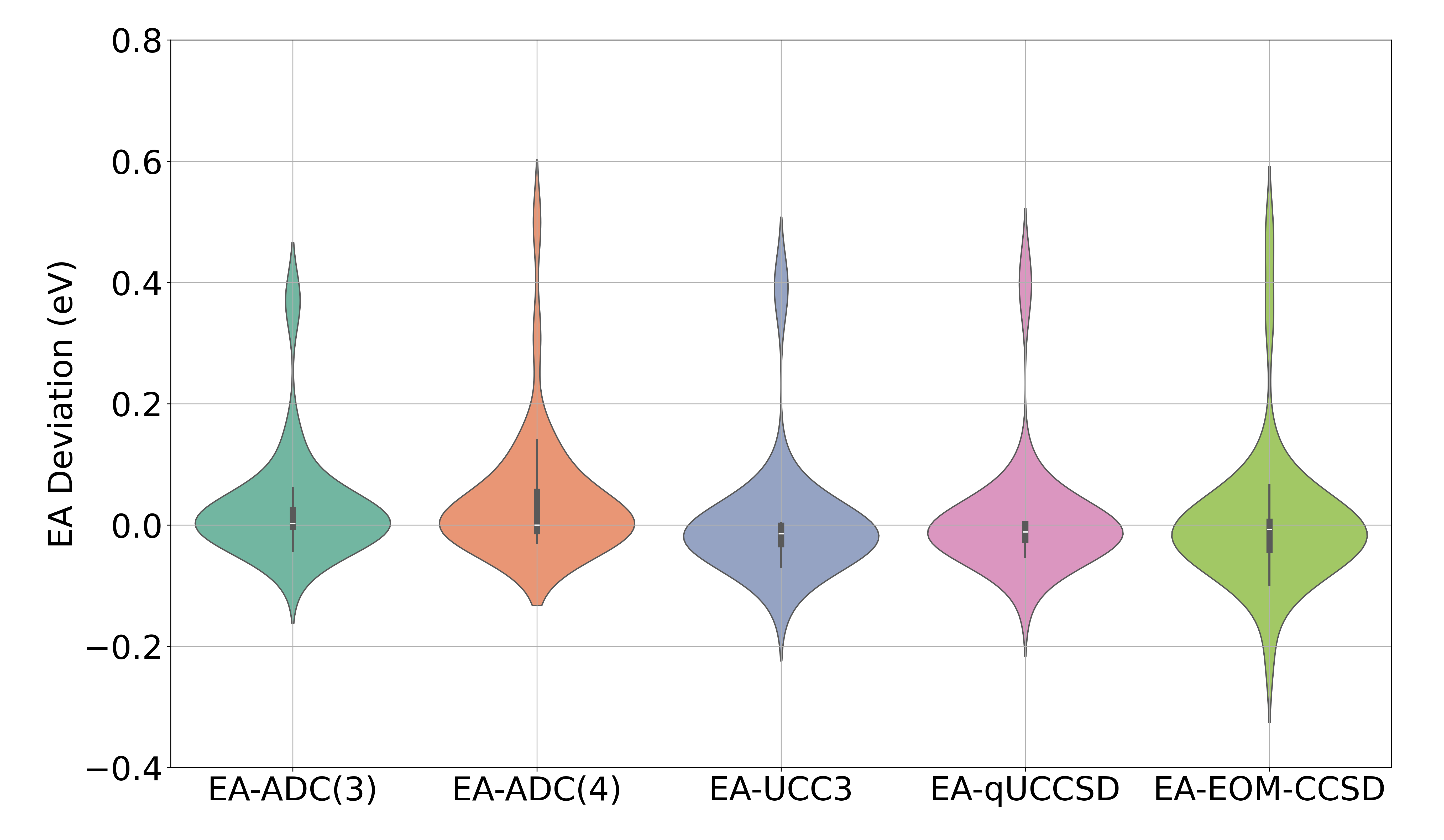}
  \caption{Violin plot of 35 1p-dominated {vertical} electron affinities obtained for systems with closed-shell reference.}
  \label{EA_violin}
\end{figure}

A statistical analysis of deviations in 35 {VEAs} with closed-shell reference
states and 16 {VEAs} with open-shell reference states, calculated using EA-qUCCSD,
EA-UCC3, EA-ADC(3), EA-ADC(4), and EA-EOM-CCSD relative to FCI reference data,
is presented in Table~\ref{t5} and Table~\ref{t6}, respectively. The complete
sets of {VEAs} computed by these methods are documented in Table~S6 and S7 of
the Supporting Information. For 1p-dominated {VEAs} with closed-shell references,
the methods show no significant difference: all yield MADs of approximately
0.05 eV and SDs around 0.10 eV. Across methods, the {VEA}
deviations fall within roughly -0.20 eV to +0.50 eV. Unlike the {VIP} case, no
clear improvement is observed when moving from ADC(3) to ADC(4) or from UCC3 to
qUCCSD; three CC-based methods display a slight tendency to underestimate {VEAs}
on average, {as reflected in their median deviations and vertical position of 
violin plots} (Figure~\ref{EA_violin}).
Over the 10 closed-shell molecules examined, the Hermitian methods perform comparably to EA-EOM-CCSD, producing
accurate and reliable electron-attachment energies. For 1p-character
{VEAs} with open-shell references, EA-ADC(4) demonstrates the highest accuracy and
precision among all methods, with a MAD of 0.03 eV and an SD of 0.06 eV -- as
expected for a higher-order approach relative to the other Hermitian methods.
Although the MAD and SD of EA-EOM-CCSD are higher than those of EA-ADC(4) by
0.05 eV and 0.06 eV, respectively, it still outperforms the other three
Hermitian methods. The performance of EA-qUCCSD closely matches that of
EA-ADC(3), consistent with trends reported for excitation-energy benchmarks of
qUCCSD and ADC(3). By contrast, EA-UCC3 is slightly worse than EA-ADC(3),
suggesting that higher-order contributions beyond third order in the double commutators -- arising from
the ground-state amplitude equations and the electron-attachment eigenvalue
equations -- tend to overcorrect the {VEAs}.

\section{Conclusion and outlook}
In this work, we derived the working equations of unitary coupled-cluster (UCC)
based electron propagator theory using self-consistent operator manifolds.
{Two} practical schemes within the UCCSD model were implemented for computing
{vertical} ionization potentials ({VIPs}) and electron affinities ({VEAs}): a
perturbation-truncated approach (IP/EA-UCC3) and a commutator-truncated
approach (IP/EA-qUCCSD). Benchmark calculations across five datasets were used
to assess the accuracy and precision of IP/EA-UCC3 and IP/EA-qUCCSD for
1h-dominated {VIPs} and 1p-dominated {VEAs}.
For systems with closed-shell reference states, IP-qUCCSD achieved the highest
accuracy among the Hermitian methods considered, outperforming both formally
third-order approaches -- IP-UCC3 and IP-ADC(3) -- and even the higher-order 
IP-ADC(4).
For open-shell systems, IP-qUCCSD performed comparably to IP-ADC(4); in both
closed- and open-shell systems, a consistent improvement from UCC3 to
qUCCSD was observed. For {VEAs} with closed-shell references, all tested methods
exhibit nearly identical performance. For the open-shell systems, EA-qUCCSD
attained accuracy similar to EA-ADC(3), reflecting trends seen for excitation
energy calculations, whereas EA-ADC(4) outperformed all other methods, as
expected for a higher-order approach. Overall, UCC-based self-consistent
electron propagator theory offers a promising framework for developing
higher-order and non-perturbative Hermitian methods for IP and EA calculations.
Future work will focus on developing and implementing advanced
commutator-truncation schemes -- such as extended qUCCSD schemes
(IP/EA-eqUCCSD)\cite{Liu2022} and cubic UCCSD schemes (IP/EA-cUCCSD))\cite{Liu2022} 
-- to further improve accuracy and explore the limits of UCCSD-based
self-consistent electron propagator theory.

\clearpage

\begin{center}
\begin{threeparttable}[!htbp]
  \caption{Components of $\bar{H}$ in the target-state eigenvalue equations for
  qUCCSD, IP-qUCCSD, and EA-qUCCSD.}
  \label{t1}
  \footnotesize
  \begin{tabular}[t]{@{}lll@{}}
  \toprule
   qUCCSD & IP-qUCCSD & EA-qUCCSD \\
  \midrule
    $\bar{H}_{\mathrm{SS}}$:       $\bar{H}_{ij}$, $\bar{H}_{ab}$, $\bar{H}_{ia,bj}$ 
   &$\bar{H}_{\mathrm{1h-1h}}$:    $\bar{H}_{ij}$
   &$\bar{H}_{\mathrm{1p-1p}}$:    $\bar{H}_{ab}$ \\
    $\bar{H}_{\mathrm{SD}}$:       $\bar{H}_{ci,ab}$, $\bar{H}_{ij,ka}$, $\bar{H}_{ibc,ajk}$ 
   &$\bar{H}_{\mathrm{1h-2h1p}}$:  $\bar{H}_{ij,ka}$
   &$\bar{H}_{\mathrm{1p-1h2p}}$:  $\bar{H}_{ci,ab}$ \\
    $\bar{H}_{\mathrm{DS}}$:       $\bar{H}_{ab,ci}$, $\bar{H}_{ka,ij}$, $\bar{H}_{ajk,ibc}$ 
   &$\bar{H}_{\mathrm{2h1p-1h}}$:  $\bar{H}_{ka,ij}$
   &$\bar{H}_{\mathrm{1h2p-1p}}$:  $\bar{H}_{ab,ci}$ \\
   $\bar{H}_{\mathrm{DD}}$:        $\bar{H}_{ij}$, $\bar{H}_{ab}$, $\bar{H}_{ia,bj}$, $\bar{H}_{ij,kl}$, $\bar{H}_{ab,cd}$ 
   &$\bar{H}_{\mathrm{2h1p-2h1p}}$:$\bar{H}_{ij}$, $\bar{H}_{ab}$, $\bar{H}_{ia,bj}$, $\bar{H}_{ij,kl}$
   &$\bar{H}_{\mathrm{1h2p-1h2p}}$:$\bar{H}_{ij}$, $\bar{H}_{ab}$, $\bar{H}_{ia,bj}$, $\bar{H}_{ab,cd}$\\
  \bottomrule
  \end{tabular}
\end{threeparttable}
\end{center}

\vspace{2em}
\begin{center}
\begin{threeparttable}[!htbp]
  \caption{Statistical deviations (eV) for 25 1h-dominated {vertical} ionization potentials
    computed with IP-EOM-CCSD, IP-ADC(3), IP-ADC(4), IP-UCC3, and IP-qUCCSD
    relative to FCI reference data\tnote{a}. All systems have closed-shell
    reference states.}
  \label{t2}
  \tabcolsep=20pt
  \begin{tabular}[t]{@{}lccccc@{}}
    \toprule
    Method   &MD &MAD &SD &MaxD(+) &MinD(-)  \\
    \midrule
    {IP-EOM-CCSD}         &0.03  &0.11  &0.13  &0.27  &-0.24  \\
    {IP-ADC(3)}           &0.26  &0.31  &0.29  &0.84  &-0.19  \\
    {IP-ADC(4)}\tnote{b}  &-0.25 &0.27  &0.27  &0.09  &-0.84  \\
    {IP-UCC3  }           &0.27  &0.27  &0.18  &0.61  &-0.05  \\
    {IP-qUCCSD}           &0.18  &0.19  &0.13  &0.46  &-0.09  \\
    \bottomrule
  \end{tabular}
  \begin{tablenotes}
  \item[a] ``MD'', ``MAD'', ``SD'', ``MaxD(+)'', and ``MinD(-)'' denote the mean
    deviation, mean absolute deviation, standard deviation, maximum positive
    deviation, and minimum negative deviation, respectively, relative to the
    FCI or CISDTQ reference values reported in ref.~\citenum{Dempwolff2020-IP-II}.
  \item[b] Ref.~\citenum{Leitner2024}.
\end{tablenotes}
\end{threeparttable}
\end{center}

\begin{center}
\begin{threeparttable}[!htbp]
  \caption{Statistical deviations (eV) for 201 1h-dominated {vertical} ionization potentials
  computed with IP-EOM-CCSD, IP-ADC(3), IP-ADC(4), IP-UCC3, and IP-qUCCSD
  relative to IP-EOM-CCSDT reference values in the CGB dataset\tnote{a}.
  All systems have closed-shell reference states.}
  \label{t3}
  \tabcolsep=20pt
  \begin{tabular}[t]{@{}lcccccc@{}}
  \toprule 
  Method   &MD &MAD &SD &MaxD(+) &MinD(-)  \\
  \midrule
  {IP-EOM-CCSD}         &0.19  &0.24  &0.28  &0.89  &-2.52 \\
  {IP-ADC(3)}           &0.27  &0.35  &0.35  &1.29  &-0.83 \\
  {IP-ADC(4)}\tnote{b}  &-0.25 &0.27  &0.27  &0.17  &-1.44 \\
  {IP-UCC3  }           &0.26  &0.30  &0.27  &0.99  &-0.83 \\
  {IP-qUCCSD}           &0.18  &0.23  &0.22  &1.00  &-0.63 \\
  \bottomrule 
  \end{tabular} 
  \begin{tablenotes}
    \item[a] ``MD'', ``MAD'', ``SD'', ``MaxD(+)'', and ``MinD(-)'' denote the mean
      deviation, mean absolute deviation, standard deviation, maximum positive
      deviation, and minimum negative deviation, resepctively, relative to the
      IP-EOM-CCSDT reference values reported in ref.~\citenum{Ranasinghe2019}.
    \item[b] Ref.~\citenum{Leitner2024}.
  \end{tablenotes}
\end{threeparttable}
\end{center}

\begin{center}
\begin{threeparttable}[!htbp]
  \caption{Statistical deviations (eV) for 17 1h-dominated {vertical} ionization potentials
    computed with IP-EOM-CCSD, IP-ADC(3), IP-ADC(4), IP-UCC3, and IP-qUCCSD
    relative to FCI reference data\tnote{a}. All systems have open-shell
    reference states.}
  \label{t4}
  \tabcolsep=20pt
  \begin{tabular}[t]{@{}lcccccc@{}}
  \toprule
  Method   &MD &MAD &SD &MaxD(+) &MinD(-)  \\
  \midrule
  {IP-EOM-CCSD}          &0.04  &0.16  &0.22  &0.34  &-0.63 \\
  {IP-ADC(3)}            &0.12  &0.24  &0.34  &0.91  &-0.63 \\
  {IP-ADC(4)}\tnote{b}   &-0.16 &0.16  &0.20  &0.03  &-0.63 \\
  {IP-UCC3  }            &0.11  &0.20  &0.26  &0.58  &-0.54 \\
  {IP-qUCCSD}            &0.06  &0.18  &0.23  &0.44  &-0.58 \\
  \bottomrule
  \end{tabular}
  \begin{tablenotes}
    \item[a] ``MD'', ``MAD'', ``SD'', ``MaxD(+)'', and ``MinD(-)'' denote the mean
    deviation, mean absolute deviation, standard deviation, maximum positive
    deviation and minimum negative deviation, respectively, relative to the
    FCI or CISDTQ reference values reported in ref.~\citenum{Dempwolff2020-IP-II}.
    \item[b] Ref.~\citenum{Leitner2024}.
  \end{tablenotes}
\end{threeparttable}
\end{center}

\begin{center}
\begin{threeparttable}[!htbp]
  \caption{Statistical deviations (eV) for 35 1p-dominated {vertical} electron affinities computed with
    {EA-EOM-CCSD, EA-ADC(3), EA-ADC(4), EA-UCC3, and EA-qUCCSD} relative to FCI
  reference data\tnote{a}. All systems have closed-shell reference states.}
  \label{t5}
  \tabcolsep=20pt
  \begin{tabular}[t]{@{}lccccc@{}}
    \toprule
    Method &MD &MAD &SD &MaxD(+) &MinD(-)  \\
    \midrule
    {EA-EOM-CCSD}         &0.01  &0.06  &0.11  &0.48  &-0.21  \\
    {EA-ADC(3)}           &0.03  &0.05  &0.09  &0.37  &-0.07  \\
    {EA-ADC(4)}\tnote{b}  &0.04  &0.05  &0.10  &0.50  &-0.03  \\
    {EA-UCC3  }           &0.01  &0.05  &0.10  &0.41  &-0.12  \\
    {EA-qUCCSD}           &0.01  &0.05  &0.10  &0.42  &-0.12  \\
    \bottomrule
  \end{tabular}
  \begin{tablenotes}
  \item[a] ``MD'', ``MAD'', ``SD'', ``MaxD(+)'', and ``MinD(-)'' denote the mean
    deviation, mean absolute deviation, standard deviation, maximum positive
    deviation and minimum negative deviation, respectively, relative to the
    FCI or CISDTQ reference values reported in ref.~\citenum{Dempwolff2021-EA}.
  \item[b] Ref.~\citenum{Leitner2024}.
\end{tablenotes}
\end{threeparttable}
\end{center}

\begin{center}
\begin{threeparttable}[!htbp] 
  \caption{Statistical deviations (eV) for 16 1p-dominated {vertical} electron affinities computed with
    {EA-EOM-CCSD, EA-ADC(3), EA-ADC(4), EA-UCC3, and EA-qUCCSD} relative to FCI
  reference data\tnote{a}. All systems have open-shell reference states.}
  \label{t6}
  \tabcolsep=20pt
  \begin{tabular}[t]{@{}lccccc@{}}
    \toprule 
    Method   &MD &MAD &SD &MaxD(+) &MinD(-)  \\
    \midrule
    {EA-EOM-CCSD}        &-0.04 &0.08  &0.12  &0.24  &-0.29  \\
    {EA-ADC(3)}          &-0.13 &0.17  &0.22  &0.16  &-0.59  \\
    {EA-ADC(4)}\tnote{b} &0.01  &0.03  &0.06  &0.16  &-0.15  \\
    {EA-UCC3  }          &-0.15 &0.20  &0.28  &0.36  &-0.79  \\
    {EA-qUCCSD}          &-0.13 &0.17  &0.24  &0.30  &-0.68  \\
    \bottomrule
  \end{tabular} 
  \begin{tablenotes} 
  \item[a] ``MD'', ``MAD'', ``SD'', ``MaxD(+)'', and ``MinD(-)'' denote the mean
     deviation, mean absolute deviation, standard deviation, maximum positive
     deviation and minimum negative deviation, respectively, relative to the
     FCI or CISDTQ reference values reported in Ref.~\citenum{Dempwolff2021-EA}.
  \item[b] Ref.~\citenum{Leitner2024}.
\end{tablenotes}
\end{threeparttable}
\end{center}

\clearpage
\appendix
\setcounter{equation}{0}
\renewcommand{\theequation}{\thesection\arabic{equation}}
\section*{Appendix: The ground-state energy expressions and amplitude equations
of the qUCCSD method}
In the qUCCSD method, the ground-state energy
$E_{\mathrm{gr}}^{\mathrm{qUCCSD}}$
includes fully contracted terms of $\bar{H}$ up to the third commutators and
can be expressed as follows:
\allowdisplaybreaks
\begin{align}
  E_{\mathrm{gr}}^{\mathrm{qUCCSD}} &= E_\mathrm{HF} + \langle \Phi_0|\bar{H}^{1}|\Phi_0\rangle
      + \langle \Phi_0|\bar{H}^{2}|\Phi_0\rangle + \langle \Phi_0|\bar{H}^{3}|\Phi_0\rangle, \\
    \langle\Phi_0|\bar{H}^{1}|\Phi_0\rangle
    &= \sum_{ijab} \frac{1}{8}  \langle ij||ab \rangle \sigma_{ij}^{ab} + h.c., \label{en_1}  \\
    \langle\Phi_0|\bar{H}^{2}|\Phi_0\rangle
    &= \sum_{ijab} \frac{1}{12} \langle ij||ab \rangle \sigma_{i}^{a} \sigma_{j}^{b} + h.c., \label{en_2}\\
    \langle\Phi_0|\bar{H}^{3}|\Phi_0\rangle &= \Bigg(
\Big( - \sum_{ijklabcd} \frac{1}{12} (\sigma_{kl}^{cd})^* \langle ij||ab \rangle 
        \sigma_{ik}^{ac} \sigma_{jl}^{bd}
      + \sum_{ijklabcd} \frac{1}{24} (\sigma_{kl}^{cd})^* \langle ij||ab \rangle 
        \sigma_{ij}^{ac} \sigma_{kl}^{bd} \nonumber \\
& \phantom{+\Bigg(}
      + \sum_{ijklabcd} \frac{1}{24} (\sigma_{kl}^{cd})^* \langle ij||ab \rangle 
        \sigma_{ik}^{ab} \sigma_{jl}^{cd}
      - \sum_{ijklabcd} \frac{1}{96} (\sigma_{kl}^{cd})^* \langle ij||ab \rangle 
        \sigma_{ij}^{cd} \sigma_{kl}^{ab} \Big) + h.c.\Bigg) \nonumber \\  
&+ \Bigg(\Big( 
      \sum_{ijklabc} \frac{1}{4} (\sigma_{jl}^{bc})^* \langle ij||ak \rangle 
      \sigma_i^a\sigma_{kl}^{bc}
    - \sum_{ijkabcd} \frac{1}{4} (\sigma_{jk}^{cd})^* \langle ic||ab \rangle 
      \sigma_i^a\sigma_{jk}^{bd} \nonumber \\
& \phantom{+\Bigg(}
    + \sum_{ijklabc} \frac{1}{2}  (\sigma_{il}^{bc})^* \langle kj||ai \rangle 
      \sigma_{j}^{b}\sigma_{lk}^{ca}
    - \sum_{ijkabcd} \frac{1}{2}  (\sigma_{jk}^{cd})^* \langle ic||ab \rangle 
      \sigma_{j}^{b}\sigma_{ik}^{ad} \nonumber \\
& \phantom{+\Bigg(}
    + \sum_{ijklabc} \frac{1}{12} (\sigma_{il}^{bc})^* \langle jk||ia \rangle \sigma_{l}^{c}\sigma_{kj}^{ab}
    - \sum_{ijkabcd} \frac{1}{12} (\sigma_{jk}^{cd})^* \langle ic||ab \rangle \sigma_{k}^{d}\sigma_{ij}^{ab} \nonumber \\
& \phantom{+\Bigg(}
    - \sum_{ijklabc} \frac{1}{8}  (\sigma_{il}^{cb})^* \langle kj||ia \rangle \sigma_{l}^{a}\sigma_{kj}^{cb}
    + \sum_{ijkabcd} \frac{1}{8}  (\sigma_{jk}^{dc})^* \langle ci||ab \rangle
      \sigma_{i}^{d}\sigma_{kj}^{ab} \Big) + h.c.\Bigg) \nonumber \\ 
&+ \Bigg(\Big(
    - \sum_{ijkabc} \frac{1}{12} (\sigma_{k}^{c})^* \langle ij||ab \rangle \sigma_i^a\sigma_{jk}^{bc}
    + \sum_{ijkabc} \frac{1}{12} (\sigma_{k}^{c})^* \langle ij||ab \rangle \sigma_i^c\sigma_{jk}^{ba}     \nonumber \\
& \phantom{+\Bigg(}
    + \sum_{ijkabc} \frac{1}{12} (\sigma_{k}^{c})^* \langle ij||ab \rangle \sigma_{k}^{a}\sigma_{ij}^{cb}
    + \sum_{ijkabc} \frac{1}{3}  (\sigma_{ik}^{bc})^* \langle jb||ai \rangle \sigma_{j}^{c}\sigma_{k}^{a} \nonumber \\
& \phantom{+\Bigg(}
    - \sum_{ijklab} \frac{1}{12} (\sigma_{ij}^{ab})^* \langle kl||ij \rangle \sigma_{k}^{a}\sigma_{l}^{b}
    - \sum_{ijabcd} \frac{1}{12} (\sigma_{ij}^{cd})^* \langle cd||ab \rangle
      \sigma_{j}^{b}\sigma_{i}^{a} \Big)  + h.c.\Bigg) \nonumber \\ 
&+ \Bigg(\Big(
      \sum_{ijkab}  \frac{1}{3}  (\sigma_{k}^{b})^* \langle ij||ak \rangle \sigma_{i}^{a}\sigma_{j}^{b}
    - \sum_{ijabc}  \frac{1}{3}  (\sigma_{j}^{a})^* \langle ai||bc \rangle
      \sigma_{j}^{b}\sigma_{i}^{c} \Big)  + h.c.\Bigg). 
\end{align}

The qUCCSD ground-state amplitude equations are equivalent to $\bar{H}_{ai} = 0 $
and $\bar{H}_{ab,ij} = 0$. Accordingly, $\bar{H}_{ai}$ consists of all
contributions from the single-commutator and double-commutator terms, 
\begin{equation}
	\bar{H}_{ai}^{\mathrm{qUCCSD}} = \bar{H}_{ai}^1 + \bar{H}_{ai}^2 = 0. 
\end{equation}
\begin{align} 
  \bar{H}_{ai}^{1} &= \sum_b f_{ab}t_i^b - \sum_j t_j^a f_{ji} +
  \frac{1}{2}\sum_{jbc}\langle aj||cb\rangle \sigma_{ij}^{cb}-
  \frac{1}{2}\sum_{jkb}\langle kj||ib\rangle \sigma_{jk}^{ba}+
  \sum_{jb}\langle aj||ib\rangle \sigma_j^{b}+
  \frac{1}{2}\sum_{jb}(\sigma_j^b)^*\langle ab||ij\rangle, \label{hbarai1} \\
  \bar{H}_{ai}^{2} &= - \sum_{jklbc} \frac{1}{2} (\sigma_{jk}^{bc})^* \langle
  al||ik\rangle  \sigma_{jl}^{bc} + \sum_{jkbcd} \frac{1}{2}
  (\sigma_{jk}^{bd})^* \langle ad||ic\rangle  \sigma_{jk}^{bc} - \sum_{jklbc}
  (\sigma_{jk}^{bc})^* \langle bl||ji\rangle  \sigma_{kl}^{ca}  \nonumber \\
  &\phantom{=} 
    +\sum_{jkbcd}(\sigma_{jk}^{bc})^* \langle ab||dj\rangle  \sigma_{ki}^{cd} 
  -\sum_{jklbc} \frac{1}{4} (\sigma_{jk}^{bc})^* \langle bl||jk\rangle
  \sigma_{il}^{ac} + \sum_{jkbcd} \frac{1}{4} (\sigma_{jk}^{bd})^* \langle
  bd||jc\rangle  \sigma_{ik}^{ac} \nonumber \\
  &\phantom{=} 
  +\sum_{jklbc} \frac{1}{4}(\sigma_{jk}^{bd})^* \langle bd||ic\rangle \sigma_{jk}^{ca} 
  -\sum_{jkbcd} \frac{1}{4}(\sigma_{jk}^{bc})^* \langle al||jk\rangle \sigma_{il}^{cb}
  +\sum_{jkbc} \frac{5}{12} \langle jk||bc\rangle \sigma_j^b \sigma_{ik}^{ac} \nonumber \\
  &\phantom{=} 
    -\sum_{jkbc} \frac{1}{3}  \langle jk||bc\rangle \sigma_k^a \sigma_{ij}^{cb}
  -\sum_{jkbc} \frac{1}{3}  \langle jk||bc\rangle \sigma_i^c \sigma_{jk}^{ba}
  -\sum_{jkbc} \frac{1}{2}  (\sigma_k^c)^* \langle cj||ib\rangle \sigma_{jk}^{ba} \nonumber \\ 
  &\phantom{=}
    -\sum_{jkbc} \frac{1}{2} (\sigma_k^c)^* \langle aj||kb\rangle \sigma_{ij}^{cb}
  -\sum_{jkbc} \frac{1}{3}  (\sigma_{jk}^{cb})^* \langle ab||ij\rangle \sigma_k^c
  -\sum_{jkbc} \frac{1}{6}  (\sigma_{jk}^{bc})^* \langle bc||ji\rangle \sigma_k^a  \nonumber \\
  &\phantom{=}
  -\sum_{jkbc} \frac{1}{6} (\sigma_{jk}^{bc})^* \langle ab||kj\rangle \sigma_i^c
  +\sum_{jbcd} \frac{1}{4} (\sigma_j^c)^* \langle ac || bd \rangle \sigma_{ij}^{bd}
  +\sum_{jklb} \frac{1}{4} (\sigma_k^b)^* \langle jl || ik \rangle \sigma_{jl}^{ab} \nonumber \\
  &\phantom{=}
  +\sum_{jbc} \langle aj||cb\rangle \sigma_j^{b} \sigma_i^c
  -\sum_{jkb}\langle kj||ib\rangle \sigma_j^{b} \sigma_k^a
  +\sum_{jbc}\frac{1}{2} (\sigma_j^b)^* \langle ab||cj\rangle \sigma_i^c
  -\sum_{jkb}\frac{1}{2} (\sigma_j^b)^* \langle kb||ij\rangle \sigma_k^a \nonumber \\ 
  &\phantom{=}
  +\sum_{jbc}\frac{1}{2} (\sigma_j^c)^* \langle ac||ib\rangle \sigma_j^b
  -\sum_{jkb}\frac{1}{2} (\sigma_j^b)^* \langle ak||ij\rangle \sigma_k^b. \label{hbarai2} 
\end{align}
The $\bar{H}_{ab,ij}$ contains all terms from the bare Hamiltonian up to
the double-commutator,
\begin{equation}
 \bar{H}_{ab,ij}^{\mathrm{qUCCSD}} = \bar{H}_{ab,ij}^{0} + \bar{H}_{ab,ij}^{1} + \bar{H}_{ab,ij}^{2} = 0.
\end{equation}
\begin{align}
  \bar{H}_{ab,ij}^{0}&=\langle ab || ij \rangle, \label{hbarabij0} \\
  \bar{H}_{ab,ij}^{1}
  &= \sum_c f_{ac} \sigma_{ij}^{cb} - \sum_k f_{ki} \sigma_{kj}^{ab}
    +\sum_{kl} \frac{1}{2} \langle kl||ij \rangle \sigma_{kl}^{ab}
    +\sum_{cd} \frac{1}{2} \langle ab||cd \rangle \sigma_{ij}^{cd}
    +P(ij) P(ab) \sum_{kc} \langle ak||ic \rangle \sigma_{jk}^{bc}\nonumber \\
  &\phantom{=}
    -P(ab) \sum_{k} \langle ka||ji\rangle \sigma_k^{b}
    +P(ij) \sum_{c} \langle ab||ic\rangle \sigma_j^{c}, \label{hbarabij1} \\
  \bar{H}_{ab,ij}^{2}
  &=
  P(ij) P(ab) \sum_{klcd} \frac{1}{3} \langle kl||cd\rangle \sigma_{ik}^{ac} \sigma_{jl}^{bd} +
  \sum_{klcd}  \frac{1}{6}  \langle kl||cd \rangle \sigma_{ij}^{cd} \sigma_{kl}^{ab} 
  -P(ab) \sum_{klcd} \frac{1}{3} \langle kl||cd \rangle \sigma_{ij}^{ad} \sigma_{kl}^{cb} \nonumber \\
  &\phantom{=}
  -P(ij) \sum_{klcd} \frac{1}{3} \langle kl||cd \rangle \sigma_{il}^{ab} \sigma_{jk}^{dc} 
  +P(ij) P(ab) \sum_{klcd}  \frac{1}{3} (\sigma_{kl}^{cd})^* \langle ad||il\rangle \sigma_{jk}^{bc} 
  +\sum_{klcd} \frac{1}{12} (\sigma_{kl}^{cd})^* \langle cd||ij\rangle \sigma_{kl}^{ab}  \nonumber \\
  &\phantom{=}
  +\sum_{klcd}  \frac{1}{12} (\sigma_{kl}^{cd})^* \langle ab||kl\rangle \sigma_{ij}^{cd} 
  -P(ab) \sum_{klcd} \frac{1}{6} (\sigma_{kl}^{cd})^* \langle ad||ij\rangle \sigma_{kl}^{cb} 
    -P(ij) \sum_{klcd} \frac{1}{6} (\sigma_{kl}^{cd})^* \langle ab||il\rangle \sigma_{jk}^{dc} \nonumber \\
  &\phantom{=}
  -P(ab) \sum_{klcd} \frac{1}{6} (\sigma_{kl}^{cd})^* \langle cb||kl\rangle \sigma_{ij}^{ad} 
  -P(ij) \sum_{klcd} \frac{1}{6} (\sigma_{kl}^{cd})^* \langle cd||kj\rangle \sigma_{il}^{ab}
  -P(ij) \sum_{klc} (\sigma_l^{c})^* \langle ck||lj\rangle \sigma_{ik}^{ab} \nonumber \\ 
  &\phantom{=}
  +P(ab) \sum_{lcd} (\sigma_l^{c})^* \langle bc||dl\rangle \sigma_{ij}^{ad}
  +P(ij) \sum_{lcd} \frac{1}{2}(\sigma_l^{c})^* \langle ab||id\rangle \sigma_{jl}^{dc}
  -P(ab) \sum_{klc} \frac{1}{2}(\sigma_l^{c})^* \langle ak||ij\rangle \sigma_{kl}^{bc} \nonumber \\ 
  &\phantom{=}
  +\sum_{klc} (\sigma_l^{c})^* \langle ck||ji\rangle  \sigma_{kl}^{ab}
  +P(ij) P(ab) \sum_{klc} (\sigma_l^{c})^* \langle bk||li\rangle \sigma_{jk}^{ca}
  -P(ij) P(ab) \sum_{lcd} (\sigma_l^{c})^* \langle ac||dj\rangle \sigma_{il}^{db} \nonumber \\ 
  &\phantom{=}
  - \sum_{lcd} (\sigma_l^{c})^* \langle ab||dl\rangle \sigma_{ij}^{dc} 
  - P(ij) \sum_{klc} \langle kl||cj\rangle \sigma_k^{c} \sigma_{il}^{ab} 
  + P(ab) \sum_{kcd}  \langle kb||cd \rangle \sigma_k^{c} \sigma_{ij}^{ad}\nonumber \\
  &\phantom{=}
  - P(ij) P(ab) \sum_{klc} \langle kl||cj \rangle \sigma_l^{b} \sigma_{ik}^{ac}
  + P(ij) P(ab) \sum_{kcd} \langle kb||cd \rangle \sigma_j^{d} \sigma_{ik}^{ac} 
  + P(ij) \sum_{klc}\frac{1}{2} \langle kl||ci\rangle \sigma_j^{c} \sigma_{kl}^{ba} \nonumber \\ 
  &\phantom{=}
  -P(ab) \sum_{kcd} \frac{1}{2} \langle ka||cd\rangle \sigma_k^{b} \sigma_{ij}^{dc}
  +P(ab) \sum_{kl} \frac{1}{2} \langle kl||ij \rangle \sigma_k^a \sigma_l^b 
  -P(ij) P(ab) \sum_{kc} \langle ak||cj \rangle \sigma_i^c \sigma_k^b \nonumber \\
  &\phantom{=}
  +P(ij) \sum_{cd} \frac{1}{2} \langle ab||cd \rangle \sigma_i^c \sigma_j^d  
  -P(ab) \sum_{kc} \frac{1}{3} (\sigma_k^{c})^* \langle ac||ij\rangle \sigma_k^b 
  -P(ij) \sum_{kc} \frac{1}{3} (\sigma_k^{c})^* \langle ab||ik\rangle \sigma_j^c.  \label{hbarabij2}
\end{align}

\begin{suppinfo}
  See the Supporting Information for computational details (molecular geometries
  and basis sets) and for complete lists of individual {VIPs} and {VEAs} for the
  five benchmark datasets computed with IP/EA-UCC3 and IP/EA-qUCCSD.
\end{suppinfo}

\section*{Note}
The authors declare no conflict of interest with regard to this content.

\begin{acknowledgement}
This work has been supported by the National Natural Science Foundation of China
(Grant No.~22173005) and Fundamental Research Funds for the Central Universities
(FRF-BR-23-02B). J. L. is grateful to Prof.~Lan Cheng at Johns Hopkins University
for helpful discussions and useful suggestions for revising the manuscript and support.
\end{acknowledgement}

\bibliography{ucc-ept}

\end{document}


\clearpage
\begin{center}
 \begin{threeparttable}[!htbp]
   \caption{Molecular geometries and basis sets used in vertical ionization potential
   (VIP) calculations with FCI reference values. All information is taken from
   ref.~\citenum{Dempwolff2020-IP-II}.}
   \tabcolsep=35pt
   \begin{tabular}[t]{@{}lll@{}}
   \toprule
     Molecule     & Geometry & Basis set  \\
   \midrule
     \ce{H2O}     &$R_{\ce{OH}}$ = 0.957 \AA,  $\angle{\ce{HOH}} = 104.5^{\circ}$   &6-31++G$^*$ \\
     \ce{NH3}     &$R_{\ce{NH}}$ = 1.0124 \AA, $\angle{\ce{HNH}} = 106.67^{\circ}$  &6-31++G$^*$ \\
     \ce{HF}      &$R_{\ce{HF}}$ = 0.917 \AA   & 6-31++G$^*$ \\
     \ce{LiH}     &$R_{\ce{LiH}}$ = 1.5957 \AA & aug-cc-pVDZ\tnote{a} \\
     \ce{CO}      &$R_{\ce{CO}}$ = 2.132 $a_0$ & cc-pVDZ \\
     \ce{HCN}     &$R_{\ce{HC}}$ = 1.064 \AA, $R_{\ce{CN}}$ = 1.156 \AA & cc-pVDZ \\
     \ce{NH^{-}2} &$R_{\ce{NH}}$ = 1.0245 \AA, $\angle{\ce{NHN}} = 103.34^{\circ}$ & aug-cc-pVDZ\tnote{b}\\
     \ce{CN^{-}}  &$R_{\ce{CN}}$ = 1.1718 \AA  &6-31+G \\
     \ce{OH^{-}}  &$R_{\ce{OH}}$ = 0.96966 \AA &cc-pVDZ \\
     \ce{NO^{-}2} &$R_{\ce{NO}}$ = 1.1934 \AA, $\angle{\ce{ONO}} = 134.1^{\circ}$ & 6-31G \\
     \ce{NO^{$\bullet$}2}   &$R_{\ce{NO}}$ = 1.1934 \AA, $\angle{\ce{ONO}} = 134.1^{\circ}$  & 6-31G \\
     \ce{NH^{$\bullet$}2}   &$R_{\ce{NH}}$ = 1.0245 \AA, $\angle{\ce{NHN}} = 103.34^{\circ}$ & aug-cc-pVDZ\tnote{b} \\
     \ce{H2O^{$\bullet$-}}  &$R_{\ce{OH}}$ = 0.957 \AA,  $\angle{\ce{HOH}} = 104.5^{\circ}$  & 6-31++G$^*$ \\
     \ce{NH3^{$\bullet$-}}  &$R_{\ce{NH}}$ = 1.0124 \AA, $\angle{\ce{HNH}} = 106.67^{\circ}$ & 6-31++G$^*$ \\
     \ce{HF^{$\bullet$-}}   &$R_{\ce{HF}}$ = 0.917 \AA   &6-31++G$^*$ \\
     \ce{LiH^{$\bullet$}-}  &$R_{\ce{LiH}}$ = 1.5957 \AA & aug-cc-pVDZ\tnote{a} \\
   \bottomrule
   \end{tabular}
   \begin{tablenotes}
   \item [a] For Li, an earlier parametrization of polarization and diffuse functions [0.00864 (s), 0.00579 (p), 0.1239 (d), 0.0725 (d)] was employed.
   \item [b] A modified aug-cc-pVDZ basis without polarization diffuse functions was used on N and H atoms.
   \end{tablenotes}
 \end{threeparttable}
\end{center}

\clearpage
\begin{longtable}{cccccccc}
  \caption{25 one-hole (1h)-dominated vertical ionization potentials (eV)
  for neutral singlet molecules and anions with closed-shell reference, computed
  using FCI, EOM-IP-CCSD, IP-ADC(3), IP-ADC(4), IP-UCC3, and IP-qUCCSD.
  To avoid redundancy, the names of the approximate methods are simplified to
  CCSD, ADC(3), ADC(4), UCC3, qUCCSD in the table header.
  Target states correspond to ($N-1$)-electron configurations relative to the
  RHF reference. For \ce{CO}, \ce{HCN}, and \ce{NO2-}, CISDTQ results are
  listed in the ``FCI'' column. FCI and IP-ADC(3) data are from
  ref.~\citenum{Dempwolff2020-IP-II}; IP-ADC(4) data are from ref.~\citenum{Leitner2024}.}
  \label{TS_IP1} \\
  \toprule
  \multicolumn{1}{c}{Molecule/target states} &\multicolumn{1}{c}{config.}  &
  \multicolumn{1}{c}{FCI}      &\multicolumn{1}{c}{CCSD}  &
  \multicolumn{1}{c}{ADC(3)}   &\multicolumn{1}{c}{ADC(4)} &
  \multicolumn{1}{c}{UCC3}     &\multicolumn{1}{c}{qUCCSD} \\

  \midrule
  \endfirsthead

  \multicolumn{2}{c}{\tablename\ \thetable{} -- {\sl (continued)}} \\
  \toprule
  \multicolumn{1}{c}{Molecule/target states} &\multicolumn{1}{c}{config.}  &
  \multicolumn{1}{c}{FCI}      &\multicolumn{1}{c}{CCSD}   &
  \multicolumn{1}{c}{ADC(3)}   &\multicolumn{1}{c}{ADC(4)} &
  \multicolumn{1}{c}{UCC3}     &\multicolumn{1}{c}{qUCCSD} \\
  \midrule
  \endhead

  \bottomrule
  \multicolumn{8}{r}{{\sl (continued)}} \\
  \endfoot

  \endlastfoot

  \ce{LiH}          &$^1\Sigma^+$       & & & & & & \\
  $1^2\Sigma^+$     &2$\sigma^{-1}$     &7.94  &7.94  &7.81  &7.90  &7.89  &7.85  \\
  \ce{NH3}          &$^1\mathrm{A}_1$   & & & & & & \\
  $1^2\mathrm{A}$   &$3a^{-1}$          &10.46 &10.44 &10.57 &10.36 &10.63 &10.60 \\
  $1^2\mathrm{E}$   &$1e^{-1}$          &16.40 &16.39 &16.50 &16.28 &16.62 &16.54 \\
  \ce{H2O}          &$^1\mathrm{A}_1$   & & & & & & \\
  $1^2\mathrm{B}_1$ &$1b_1^{-1}$        &12.30 &12.17 &12.72 &11.98 &12.57 &12.52 \\
  $1^2\mathrm{A}_1$ &$3a_1^{-1}$        &14.63 &14.50 &15.04 &14.29 &14.93 &14.84 \\
  $1^2\mathrm{B}_2$ &$1b_2^{-1}$        &18.97 &18.89 &19.30 &18.67 &19.25 &19.13 \\
  \ce{HF}           &$^1\Sigma^+$       & & & & & & \\
  $1^2\Pi$          &1$\pi^{-1}$        &15.93 &15.69 &16.68 &15.35 &16.31 &16.21 \\
  $1^2\Sigma^+$     &3$\sigma^{-1}$     &19.98 &19.78 &20.58 &19.49 &20.30 &20.16 \\
  \ce{CO}           &$^1\Sigma^+$       & & & & & & \\
  $1^2\Sigma^+$     &5$\sigma^{-1}$     &13.56 &13.83 &13.38 &13.62 &13.65 &13.62 \\
  $1^2\Pi$          &1$\pi^{-1}$        &16.68 &16.75 &16.88 &16.57 &16.68 &16.63 \\
  $2^2\Sigma^+$     &4$\sigma^{-1}$     &19.39 &19.49 &20.23 &18.59 &19.88 &19.68 \\
  \ce{HCN}          &$^1\Sigma^+$       & & & & & & \\
  $1^2\Pi$          &1$\pi^{-1}$        &13.45 &13.66 &13.26 &13.53 &13.46 &13.44 \\
  $1^2\Sigma^+$     &5$\sigma^{-1}$     &13.59 &13.63 &13.90 &13.44 &13.89 &13.86 \\
  $2^2\Sigma^+$     &4$\sigma^{-1}$     &20.14 &20.37 &20.22 &20.15 &20.30 &20.28 \\
  \ce{NH2-}         &$^1\mathrm{A}_1$   & & & & & & \\
  $1^2\mathrm{B}_1$ &1$b_1^{-1}$        &0.31  &0.35  &0.70  &-0.18 &0.55  &0.43  \\
  $1^2\mathrm{A}_1$ &3$a_1^{-1}$        &2.49  &2.55  &3.09  &1.75  &3.04  &2.81  \\
  $2^2\mathrm{B}_2$ &1$b_2^{-1}$        &7.02  &7.13  &7.63  &6.18  &7.63  &7.36  \\
  \ce{OH-}          &$^1\Sigma^+$       & & & & & & \\
  $1^2\Pi$          &1$\pi^{-1}$        &-0.62 &-0.73 &-0.47 &-0.68 &-0.43 &-0.45 \\
  $1^2\Sigma^+$     &3$\sigma^{-1}$     &3.74  &3.68  &3.80  &3.71  &3.85  &3.83  \\
  \ce{CN-}          &$^1\Sigma^+$       & & & & & & \\
  $1^2\Sigma^+$     &5$\sigma^{-1}$     &3.26  &3.41  &3.22  &3.23  &3.48  &3.46  \\
  $1^2\Pi$          &1$\pi^{-1}$        &5.11  &5.29  &5.02  &4.96  &5.42  &5.23  \\
  $2^2\Sigma^+$     &4$\sigma^{-1}$     &6.46  &6.60  &7.01  &6.12  &7.01  &6.92  \\
  \ce{NO2-}         &$^1\mathrm{A}_1$   & & & & & & \\
  $1^2\mathrm{A}_1$ &6$a_1^{-1}$        &-0.17 &-0.08 &-0.08 &-0.17 &0.14  &0.11  \\
  $1^2\mathrm{B}_2$ &4$b_2^{-1}$        &3.32  &3.41  &3.91  &2.90  &3.90  &3.74  \\
  $1^2\mathrm{A}_2$ &1$a_2^{-1}$        &3.58  &3.59  &3.61  &3.67  &3.65  &3.62  \\
  \bottomrule
\end{longtable}

\clearpage
\begin{center}
  \small
  \begin{longtable}{cccccccc}
    \caption{17 one-hole (1h)-dominated vertical ionization potentials (eV)
    for radical, neutral triplet molecules, and radical anions with open-shell
    reference, computed using FCI, EOM-IP-CCSD, IP-ADC(3), IP-ADC(4),
    IP-UCC3, and IP-qUCCSD methods.
    To avoid redundancy, the names of the approximate methods are simplified to
    CCSD, ADC(3), ADC(4), UCC3, qUCCSD in the table header.
    Target states correspond to ($N-1$)- electron configurations relative to
    the UHF reference. For \ce{NO2^{$\bullet$}} and \ce{HCN^{$\bullet$-}},
    CISDTQ results are shown in the ``FCI'' column.
    FCI and IP-ADC(3) data are from ref.~\citenum{Dempwolff2020-IP-II};
    IP-ADC(4) data are from ref.~\citenum{Leitner2024}.}
    \label{TS_IP2} \\
    \toprule
    \multicolumn{1}{c}{Molecule/target states} &\multicolumn{1}{c}{config.}  &
    \multicolumn{1}{c}{FCI}      &\multicolumn{1}{c}{CCSD}   &
    \multicolumn{1}{c}{ADC(3)}   &\multicolumn{1}{c}{ADC(4)} &
    \multicolumn{1}{c}{UCC3}     &\multicolumn{1}{c}{qUCCSD} \\

    \midrule
    \endfirsthead

    \multicolumn{2}{c}{\tablename\ \thetable{} -- {\sl (continued)}} \\
    \toprule
    \multicolumn{1}{c}{Molecule/target states} &\multicolumn{1}{c}{config.}  &
    \multicolumn{1}{c}{FCI}      &\multicolumn{1}{c}{CCSD}   &
    \multicolumn{1}{c}{ADC(3)}   &\multicolumn{1}{c}{ADC(4)} &
    \multicolumn{1}{c}{UCC3}     &\multicolumn{1}{c}{qUCCSD} \\
    \midrule
    \endhead

    \bottomrule
    \multicolumn{8}{r}{{\sl (continued)}} \\
    \endfoot

    \endlastfoot

    \ce{NH_2^{$\bullet$}} &$^2\mathrm{B}_1$  & & & & & & \\
     $1^3\mathrm{B}_1$    &3$a_1^{-1}$       &11.68 &11.63 &11.75 &11.57 &11.81 &11.75 \\
     $1^1\mathrm{A}_1$    &1$b_1^{-1}$       &12.20 &12.21 &12.34 &12.11 &12.35 &12.35 \\
     \ce{NO2^{$\bullet$}} &$^2\mathrm{A}_1$  & & & & & & \\
     $1^1\mathrm{A}_1$    &$6a_1^{-1}$       &10.84 &10.80 &10.78 &10.35 &11.10 &11.06 \\
     $1^3\mathrm{B}_2$    &$4b_2^{-1}$       &13.19 &12.56 &12.56 &12.66 &12.65 &12.61 \\
     \ce{NH3^{*}}         &$^3\mathrm{A}(3a \rightarrow 4a)$   & & & & & & \\
     $1^2\mathrm{A}$      &$4a^{-1}$         &4.46  &4.45  &4.43  &4.44  &4.46  &4.43  \\
     $2^2\mathrm{A}$      &$3a^{-1}$         &17.54 &17.87 &17.94 &17.57 &17.88 &17.86 \\
     \ce{H2O^{*}}         &$^3\mathrm{B}_1(1b_1 \rightarrow 4a_1)$   & & & & & & \\
     $1^2\mathrm{B}_1$    &$4a_1^{-1}$       &5.30  &5.28  &5.28  &5.28  &5.29  &5.27  \\
     $1^4\mathrm{B}_1$    &$3a_1^{-1}$       &18.68 &18.90 &19.02 &18.52 &18.91 &18.84 \\
     $2^2\mathrm{A}_1$    &$1b_1^{-1}$       &19.99 &20.33 &20.61 &19.91 &20.46 &20.43 \\
     \ce{LiH^{$\bullet$-}}&$^2\Sigma^+$      & & & & & & \\
     $1^1\Sigma^+$        &3$\sigma^{-1}$    &0.30  &0.30  &0.29  &0.30  &0.30  &0.30  \\
     $1^3\Sigma^+$        &2$\sigma^{-1}$    &3.49  &3.75  &3.41  &3.49  &3.43  &3.36  \\
     \ce{NH3^{$\bullet$-}}&$^2\mathrm{A}$    & & & & & & \\
     $1^1\mathrm{A}$      &4$a^{-1}$         &-0.95 &-0.97 &-0.98 &-0.97 &-0.96 &-0.98 \\
     $1^3\mathrm{A}$      &3$a^{-1}$         &5.04  &5.28  &5.35  &4.87  &5.38  &5.27  \\
     \ce{H2O^{$\bullet$-}}&$^2\mathrm{A}_2$  & & & & & & \\
     $1^1\mathrm{A}_1$    &$4a_1^{-1}$       &-0.94 &-0.95 &-0.96 &-0.96 &-0.94 &-0.96 \\
     $1^3\mathrm{B}_1$    &$1b_1^{-1}$       &6.06  &6.30  &6.70  &5.69  &6.53  &6.37  \\
     \ce{HF^{$\bullet$-}} &$^2\Sigma^+$      & & & & & & \\
     $1^1\Sigma^+$        &4$\sigma^{-1}$    &-0.99 &-1.01 &-1.00 &-1.01 &-0.99 &-1.01 \\
     $1^3\Pi$             &1$\pi^{-1}$       &9.00  &9.25  &9.91  &8.37  &9.58  &9.37  \\
     \bottomrule
  \end{longtable}
\end{center}

\clearpage
\begin{center}
  \begin{longtable}{ccccccccc}
  \caption{201 1h-dominated vertical ionization potentials (eV) for neutral
    molecules with closed-shell reference states, computed using EOM-IP-CCSDTQ,
    EOM-IP-CCSDT, EOM-IP-CCSD, IP-ADC(3), IP-ADC(4), IP-UCC3, and IP-qUCCSD methods. 
    To avoid redundancy, the names of the approximate methods are simplified to
    CCSD, ADC(3), ADC(4), UCC3, qUCCSD in the table header.
    Target states correspond to ($N-1$)-electron configurations
    relative to the RHF reference. The cc-pVTZ basis set was used for all
    calculations. Molecular geometries and IP-EOM-CCSDTQ/IP-EOM-CCSDT results
    are from ref.~\citenum{Ranasinghe2019}; IP-ADC(3) and IP-ADC(4) results
    are from ref.~\citenum{Leitner2024}.}
  \label{TS_IP3} \\

  \toprule
  \multicolumn{1}{c}{Molecule} &\multicolumn{1}{c}{config.} &
  \multicolumn{1}{c}{CCSDTQ}   &\multicolumn{1}{c}{CCSDT}   &
  \multicolumn{1}{c}{CCSD}     &\multicolumn{1}{c}{ADC(3)}  &
  \multicolumn{1}{c}{ADC(4)}   &\multicolumn{1}{c}{UCC3}    &
  \multicolumn{1}{c}{qUCCSD} \\

  \midrule
  \endfirsthead

  \multicolumn{2}{c}{\tablename\ \thetable{} -- {\sl (continued)}} \\
  \toprule
  \multicolumn{1}{c}{Molecule} &\multicolumn{1}{c}{config.} &
  \multicolumn{1}{c}{CCSDTQ}   &\multicolumn{1}{c}{CCSDT}   &
  \multicolumn{1}{c}{CCSD}     &\multicolumn{1}{c}{ADC(3)}  &
  \multicolumn{1}{c}{ADC(4)}   &\multicolumn{1}{c}{UCC3}    &
  \multicolumn{1}{c}{qUCCSD} \\
  \midrule
  \endhead

  \bottomrule
  \multicolumn{9}{r}{{\sl (continued)}} \\
  \endfoot

  \endlastfoot

  \ce{CO}  &$5\sigma^{-1}$   &13.88 &13.90 &14.14 &13.69 &13.93 &13.94 &13.92\\
           &$1\pi^{-1}$      &16.94 &16.95 &17.01 &17.06 &16.82 &16.89 &16.84\\
           &$4\sigma^{-1}$   &19.57 &19.54 &19.74 &20.34 &18.79 &20.03 &19.84\\
 \ce{H2O}  &$1b_1^{-1}$      &12.49 &12.45 &12.40 &12.73 &12.33 &12.67 &12.65\\
           &$3a_1^{-1}$      &14.73 &14.69 &14.63 &14.94 &14.57 &14.90 &14.86\\
           &$1b_2^{-1}$      &18.86 &18.84 &18.81 &18.98 &18.75 &18.97 &18.93\\
  \ce{HF}  &$1\pi^{-1}$      &15.95 &15.90 &15.81 &16.39 &15.66 &16.23 &16.17\\
           &$3\sigma^{-1}$   &19.86 &19.83 &19.75 &20.15 &19.65 &20.02 &19.97\\
  \ce{F2}  &$1\pi_{g}^{-1}$  &15.61 &15.58 &15.51 &15.86 &15.39 &15.84 &15.74\\
           &$1\pi_{u}^{-1}$  &18.73 &18.74 &18.82 &19.04 &18.45 &18.96 &18.84\\
           &$3\sigma_g^{-1}$ &20.96 &20.96 &21.05 &20.86 &20.92 &20.98 &20.91\\
  \ce{HCl} &$2\pi^{-1}$      &12.59 &12.58 &12.63 &12.58 &12.56 &12.56 &12.57\\
           &$5\sigma^{-1}$   &16.58 &16.57 &16.64 &16.58 &16.56 &16.56 &16.57\\
           &$4\sigma^{-1}$   &26.49 &26.63 &25.69 &26.82 &26.54 &26.85 &26.84\\
  \ce{CH4} &$1t_2^{-1}$      &14.36 &14.35 &14.38 &14.32 &14.34 &14.40 &14.40\\
           &$2a_1^{-1}$      &23.12 &23.11 &23.35 &23.21 &23.10 &23.34 &23.33\\
  \ce{NH3} &$3a_1^{-1}$      &10.76 &10.74 &10.73 &10.81 &10.72 &10.85 &10.84\\
           &$1e_1^{-1}$      &16.50 &16.48 &16.48 &16.51 &16.45 &16.59 &16.56\\
  \ce{N2}  &$3\sigma_g^{-1}$ &15.42 &15.45 &15.57 &15.44 &15.47 &15.53 &15.57\\
           &$1\pi_u^{-1}$    &16.88 &16.89 &17.17 &16.53 &17.06 &16.81 &16.82\\
           &$2\sigma_u^{-1}$ &18.60 &18.67 &18.83 &18.81 &18.61 &18.82 &18.80\\
  \ce{CO2} &$1\pi_g^{-1}$    &13.64 &13.63 &13.70 &13.74 &13.50 &13.72 &13.62\\
           &$1\pi_u^{-1}$    &17.46 &17.50 &17.95 &17.74 &17.15 &17.77 &17.57\\
           &$3\sigma_u^{-1}$ &17.88 &17.84 &18.09 &18.51 &17.26 &18.43 &18.23\\
           &$4\sigma_g^{-1}$ &19.11 &19.08 &19.41 &19.90 &18.34 &19.78 &19.54\\
  \ce{C2H2}&$1\pi_u^{-1}$    &11.42 &11.42 &11.54 &11.19 &11.47 &11.33 &11.34\\
           &$3\sigma_g^{-1}$ &17.10 &17.12 &17.24 &17.18 &17.08 &17.22 &17.26\\
           &$2\sigma_u^{-1}$ &18.97 &19.00 &19.15 &19.15 &18.92 &19.18 &19.14\\
           &$2\sigma_g^{-1}$ &23.60 &23.70 &24.43 &23.80 &23.77 &23.86 &23.87\\
  \ce{HCN} &$1\pi^{-1}$      &13.71 &13.72 &13.89 &13.44 &13.79 &13.63 &13.63\\
           &$5\sigma^{-1}$   &13.83 &13.84 &13.92 &14.12 &13.68 &14.08 &14.08\\
           &$4\sigma^{-1}$   &20.36 &20.43 &20.63 &20.44 &20.38 &20.51 &20.50\\
  \ce{H2CO}&$2b_2^{-1}$      &10.79 &10.75 &10.76 &11.09 &10.43 &11.04 &10.93\\
           &$1b_1^{-1}$      &14.50 &14.50 &14.55 &14.40 &14.36 &14.38 &14.33\\
           &$5a_1^{-1}$      &16.01 &15.96 &16.03 &16.47 &15.49 &16.43 &16.25\\
           &$1b_2^{-1}$      &17.05 &17.10 &17.44 &17.15 &16.90 &17.20 &17.14\\
           &$4a_1^{-1}$      &21.24 &21.30 &21.72 &21.44 &21.08 &21.53 &21.45\\
  \ce{P2}  &$2\pi_u^{-1}$    &10.49 &10.51 &10.56 &10.29 &10.54 &10.25 &10.28\\
           &$5\sigma_g^{-1}$ &10.64 &10.65 &10.74 &10.62 &10.60 &10.56 &10.59\\
  \ce{CS}  &$7\sigma^{-1}$   &11.23 &11.26 &11.47 &11.02 &11.34 &11.26 &11.27\\
           &$2\pi^{-1}$      &12.85 &12.88 &12.96 &12.63 &13.01 &12.63 &12.64\\
           &$6\sigma^{-1}$   &17.89 &18.08 &17.17 &17.98 &17.90 &17.88 &17.77\\
  \ce{OCS} &$3\pi^{-1}$      &-     &11.12 &11.20 &10.93 &11.18 &10.99 &10.97\\
           &$2\pi^{-1}$      &-     &15.61 &16.14 &15.76 &15.24 &15.73 &15.47\\
           &$9\sigma^{-1}$   &-     &15.83 &16.10 &15.90 &15.63 &15.98 &15.93\\
           &$8\sigma^{-1}$   &-     &17.89 &18.40 &18.90 &17.01 &18.68 &18.36\\
  \ce{FCN} &$2\pi^{-1}$      &-     &13.50 &13.70 &13.26 &13.58 &13.48 &13.45\\
           &$7\sigma^{-1}$   &-     &14.31 &14.44 &14.54 &14.17 &14.56 &14.54\\
           &$1\pi^{-1}$      &-     &19.39 &19.67 &20.10 &18.89 &19.91 &19.76\\
           &$6\sigma^{-1}$   &-     &22.79 &23.07 &23.36 &22.32 &23.27 &23.10\\
  \ce{C2H6}&$1e_g^{-1}$      &-     &12.68 &12.71 &12.66 &12.66 &12.73 &12.72\\
           &$3a_{1g}^{-1}$   &-     &13.06 &13.11 &13.08 &13.03 &13.16 &13.18\\
           &$1e_u^{-1}$      &-     &15.38 &15.44 &15.40 &15.35 &15.48 &15.47\\
           &$2b_u^{-1}$      &-     &20.69 &20.90 &20.82 &20.66 &20.91 &20.89\\
           &$2a_{1g}^{-1}$   &-     &24.15 &24.60 &24.45 &24.08 &24.56 &24.54\\
  \ce{CH2F2}&$2b_1^{-1}$     &-     &13.29 &13.35 &13.54 &13.08 &13.54 &13.47\\
            &$4b_2^{-1}$     &-     &14.92 &14.95 &15.56 &14.53 &15.37 &15.25\\
            &$6a_1^{-1}$     &-     &15.20 &15.25 &15.61 &14.91 &15.51 &15.43\\
            &$1a_2^{-1}$     &-     &15.59 &15.66 &16.32 &15.15 &16.13 &15.99\\
            &$3b_2^{-1}$     &-     &18.75 &18.85 &19.19 &18.45 &19.10 &18.98\\
            &$5a_1^{-1}$     &-     &18.97 &19.13 &19.39 &18.70 &19.34 &19.24\\
            &$1b_1^{-1}$     &-     &19.15 &19.33 &19.51 &18.93 &19.51 &19.42\\
  \ce{CH3F} &$2e_1^{-1}$     &-     &13.17 &13.18 &13.48 &12.92 &13.43 &13.37\\
            &$5a_1^{-1}$     &-     &17.01 &17.04 &17.29 &16.78 &17.21 &17.15\\
            &$1e_1^{-1}$     &-     &17.04 &17.14 &17.42 &16.84 &17.36 &17.29\\
            &$4a_1^{-1}$     &-     &23.51 &23.79 &23.69 &23.41 &23.78 &23.72\\
  \ce{CHF3} &$6a_1^{-1}$     &-     &14.76 &14.87 &15.02 &14.58 &15.02 &14.95\\
            &$1a_2^{-1}$     &-     &15.33 &15.40 &15.96 &14.98 &15.80 &15.68\\
            &$5e_1^{-1}$     &-     &15.93 &16.02 &16.52 &15.58 &16.39 &16.26\\
            &$4e_1^{-1}$     &-     &16.99 &17.11 &17.59 &16.64 &17.46 &17.33\\
            &$3e_1^{-1}$     &-     &20.41 &20.59 &20.87 &20.11 &20.83 &20.69\\
            &$5a_1^{-1}$     &-     &20.93 &21.81 &21.31 &20.69 &21.34 &21.23\\
            &$4a_1^{-1}$     &-     &24.43 &24.74 &24.86 &24.16 &24.88 &24.75\\
  \ce{CH3CCH}&$2e_1^{-1}$    &-     &10.46 &10.57 &10.24 &10.50 &10.39 &10.38\\
             &$1e_1^{-1}$    &-     &15.17 &15.28 &15.21 &15.13 &15.27 &15.26\\
             &$7a_1^{-1}$    &-     &15.22 &15.35 &15.31 &15.17 &15.39 &15.41\\
             &$6a_1^{-1}$    &-     &17.65 &17.83 &17.81 &17.54 &17.85 &17.81\\
  \ce{NSF}  &$13a^{'-1}$     &-     &11.62 &11.76 &11.45 &11.54 &11.21 &11.55\\
            &$12a^{'-1}$     &-     &13.30 &13.52 &13.47 &12.91 &13.51 &13.43\\
            &$3a^{''-1}$     &-     &13.74 &13.92 &13.67 &13.32 &13.79 &13.71\\
            &$11a^{'-1}$     &-     &15.24 &15.59 &16.34 &14.36 &16.09 &15.97\\
            &$2a^{''-1}$     &-     &16.11 &16.48 &17.03 &15.52 &16.48 &16.36\\
            &$10a^{'-1}$     &-     &16.35 &16.86 &17.25 &15.71 &17.14 &16.75\\
  \ce{HCOOH}&$10a^{'-1}$     &-     &11.33 &11.40 &11.74 &10.97 &11.68 &11.53\\
            &$2a^{''-1}$     &-     &12.47 &12.52 &12.60 &12.34 &12.56 &12.47\\
            &$9a^{'-1}$      &-     &14.79 &14.92 &15.19 &14.47 &15.20 &15.05\\
            &$1a^{''-1}$     &-     &15.71 &16.02 &15.93 &15.48 &15.95 &15.78\\
            &$8a^{'-1}$      &-     &17.06 &17.32 &17.78 &16.49 &17.71 &17.48\\
            &$7a^{'-1}$      &-     &17.81 &18.18 &18.07 &17.56 &18.15 &18.03\\
            &$6a^{'-1}$      &-     &22.08 &22.57 &22.59 &21.84 &22.67 &22.55\\
  \ce{SiO}  &$7\sigma^{-1}$  &-     &11.33 &11.43 &11.70 &10.09 &11.66 &11.47\\
            &$2\pi^{-1}$     &-     &11.87 &11.92 &12.64 &10.74 &12.13 &11.88\\
            &$6\sigma^{-1}$  &-     &14.58 &15.05 &15.51 &13.43 &15.21 &14.86\\
  \ce{H2CS} &$3b_2^{-1}$     &-     & 9.25 & 9.27 & 9.22 & 9.16 & 9.29 & 9.19\\
            &$2b_1^{-1}$     &-     &11.78 &11.83 &11.52 &11.78 &11.61 &11.59\\
            &$7a_1^{-1}$     &-     &13.88 &13.99 &13.85 &13.86 &13.88 &13.88\\
            &$2b_2^{-1}$     &-     &15.61 &15.98 &15.60 &15.41 &15.61 &15.61\\
            &$6a_1^{-1}$     &-     &19.60 &19.50 &19.57 &19.36 &19.58 &19.53\\
  \ce{CF4}  &$1t_1^{-1}$     &-     &16.09 &16.21 &16.66 &15.79 &16.55 &16.43\\
            &$4t_2^{-1}$     &-     &17.24 &17.37 &17.68 &16.98 &17.63 &17.51\\
            &$1e^{-1}$       &-     &18.18 &18.36 &18.76 &17.87 &18.67 &18.55\\
            &$3t_2^{-1}$     &-     &21.93 &22.18 &22.37 &21.67 &22.37 &22.24\\
            &$4a_1^{-1}$     &-     &24.85 &25.16 &25.39 &24.54 &25.38 &25.23\\
  \ce{SiF4} &$1t_1^{-1}$     &-     &16.23 &16.37 &16.90 &15.85 &16.72 &16.58\\
            &$5t_2^{-1}$     &-     &17.24 &17.40 &17.88 &16.86 &17.72 &17.58\\
            &$1e^{-1}$       &-     &17.63 &17.80 &18.28 &17.23 &18.12 &17.99\\
            &$4t_2^{-1}$     &-     &19.17 &19.35 &19.80 &18.76 &19.66 &19.52\\
            &$5a_1^{-1}$     &-     &21.27 &21.49 &21.94 &20.83 &21.80 &21.66\\
  \ce{HCCF} &$2\pi^{-1}$     &-     &11.34 &11.49 &11.13 &11.39 &11.30 &11.29\\
            &$1\pi^{-1}$     &-     &17.90 &18.12 &18.73 &17.33 &18.47 &18.32\\
            &$7\sigma^{-1}$  &-     &18.12 &18.29 &18.19 &18.07 &18.28 &18.27\\
            &$6\sigma^{-1}$  &-     &20.95 &21.26 &21.71 &20.50 &21.59 &21.47\\
            &$5\sigma^{-1}$  &-     &24.40 &25.29 &24.51 &24.39 &24.60 &24.59\\
  \ce{NNO}  &$2\pi^{-1}$     &-     &12.74 &12.83 &12.63 &12.53 &12.73 &12.65\\
            &$7\sigma^{-1}$  &-     &16.28 &16.64 &16.28 &16.25 &16.60 &16.51\\
            &$1\pi^{-1}$     &-     &18.31 &19.00 &19.05 &17.87 &19.14 &19.07\\
            &$6\sigma^{-1}$  &-     &19.89 &20.15 &20.94 &18.45 &20.61 &20.26\\
  \ce{CH3NC}&$7a_1^{-1}$     &-     &11.21 &11.45 &11.04 &11.21 &11.24 &11.22\\
            &$2e_1^{-1}$     &-     &12.55 &12.64 &12.50 &12.53 &12.51 &12.48\\
            &$1e_1^{-1}$     &-     &16.59 &16.68 &16.81 &16.41 &16.74 &16.68\\
            &$6a_1^{-1}$     &-     &18.38 &18.61 &18.40 &18.08 &18.35 &18.33\\
            &$5a_1^{-1}$     &-     &25.06 &25.48 &25.68 &24.79 &25.64 &25.62\\
  \ce{CH3CN}&$2e_1^{-1}$     &-     &12.45 &12.60 &12.19 &12.51 &12.38 &12.35\\
            &$7a_1^{-1}$     &-     &12.97 &13.11 &13.28 &12.77 &13.27 &13.23\\
            &$1e_1^{-1}$     &-     &16.24 &16.40 &16.22 &16.23 &16.32 &16.31\\
            &$6a_1^{-1}$     &-     &17.38 &17.54 &17.42 &17.34 &17.53 &17.54\\
            &$5a_1^{-1}$     &-     &24.88 &25.36 &25.30 &24.79 &25.40 &25.39\\
\ce{gem-C2H2F2}&$2b_1^{-1}$  &-     &10.54 &10.64 &10.39 &10.54 &10.55 &10.51\\
               &$5b_2^{-1}$  &-     &14.96 &15.10 &15.17 &14.64 &15.23 &15.17\\
               &$8a_1^{-1}$  &-     &15.52 &15.69 &15.81 &15.26 &15.82 &15.76\\
               &$4b_2^{-1}$  &-     &15.85 &16.01 &16.37 &15.62 &16.23 &16.10\\
               &$1a_2^{-1}$  &-     &16.02 &16.22 &16.74 &15.58 &16.52 &16.39\\
               &$7a_1^{-1}$  &-     &18.27 &18.35 &18.51 &17.85 &18.48 &18.39\\
               &$1b_1^{-1}$  &-     &18.10 &18.39 &18.76 &17.99 &18.62 &18.48\\
               &$3b_2^{-1}$  &-     &19.63 &19.89 &20.14 &19.24 &20.05 &19.90\\
               &$6a_1^{-1}$  &-     &21.41 &21.81 &22.04 &21.00 &22.00 &21.70\\
               &$5a_1^{-1}$  &-     &25.10 &25.58 &25.36 &25.12 &25.43 &25.41\\
\ce{cis-C2H2F2}&$2b_1^{-1}$  &-     &10.39 &10.47 &10.27 &10.34 &10.41 &10.37\\
               &$7a_1^{-1}$  &-     &13.88 &14.01 &14.24 &13.55 &14.20 &14.13\\
               &$6b_2^{-1}$  &-     &14.80 &14.94 &15.32 &14.32 &15.22 &15.10\\
               &$1a_2^{-1}$  &-     &16.07 &16.22 &16.84 &15.53 &16.57 &16.44\\
               &$5b_2^{-1}$  &-     &16.85 &17.08 &17.40 &16.67 &17.31 &17.20\\
               &$1b_1^{-1}$  &-     &16.97 &17.11 &17.52 &16.38 &17.35 &17.20\\
               &$6a_1^{-1}$  &-     &18.50 &18.71 &18.97 &18.11 &18.90 &18.79\\
               &$5a_1^{-1}$  &-     &18.88 &19.17 &19.24 &18.55 &19.24 &19.15\\
               &$4b_2^{-1}$  &-     &20.85 &21.19 &21.30 &20.47 &21.26 &21.13\\
               &$4a_2^{-1}$  &-     &24.61 &25.09 &24.70 &24.62 &24.77 &24.72\\
  \ce{O3}   &$6a_1^{-1}$     &-     &12.54 &12.74 &12.72 &11.99 &12.93 &12.67\\
            &$4b_2^{-1}$     &-     &12.69 &12.87 &12.87 &12.02 &12.93 &12.69\\
            &$1a_2^{-1}$     &-     &13.46 &13.43 &12.63 &13.58 &12.63 &12.83\\
            &$1b_1^{-1}$     &-     &21.32 &18.80 &20.76 &21.20 &21.30 &20.97\\
\ce{HCONH2} &$10a^{'-1}$     &-     &10.17 &10.29 &10.60 & 9.71 &10.58 &10.40\\
            &$2a^{''-1}$     &-     &10.56 &10.61 &10.68 &10.48 &10.59 &10.52\\
            &$1a^{''-1}$     &-     &14.06 &14.44 &14.17 &13.86 &14.20 &14.00\\
            &$9a^{'-1}$      &-     &14.67 &14.88 &15.28 &14.09 &15.23 &15.00\\
            &$8a^{'-1}$      &-     &16.55 &16.90 &16.93 &16.32 &17.00 &16.90\\
            &$7a^{'-1}$      &-     &19.25 &19.06 &19.29 &18.99 &19.36 &19.36\\
            &$6a^{'-1}$      &-     &20.90 &21.26 &21.11 &20.68 &21.20 &21.13\\
  \ce{C2H4} &$1b_{3u}^{-1}$  &-     &10.64 &10.66 &10.44 &10.62 &10.59 &10.55\\
            &$1b_{3g}^{-1}$  &-     &13.08 &13.12 &13.05 &13.04 &13.11 &13.11\\
            &$3a_g^{-1}$     &-     &14.79 &14.90 &14.80 &14.77 &14.87 &14.90\\
            &$1b_{2u}^{-1}$  &-     &16.14 &16.31 &16.11 &16.05 &16.20 &16.17\\
            &$2b_{1u}^{-1}$  &-     &19.32 &19.60 &19.31 &19.22 &19.40 &19.35\\
            &$2a_g^{-1}$     &-     &23.87 &24.43 &24.06 &23.77 &24.11 &24.11\\
  \ce{C2F4} &$2b_{3u}^{-1}$  &-     &10.37 &10.51 &10.22 &10.37 &10.43 &10.39\\
            &$4b_{3g}^{-1}$  &-     &15.78 &15.97 &16.50 &15.30 &16.32 &16.18\\
            &$6a_g^{-1}$     &-     &16.11 &16.31 &16.52 &15.79 &16.49 &16.40\\
            &$4b_{2u}^{-1}$  &-     &16.32 &16.53 &17.00 &15.82 &16.84 &16.70\\
            &$1a_u^{-1}$     &-     &16.48 &16.70 &17.17 &16.02 &17.00 &16.86\\
            &$1b_{1g}^{-1}$  &-     &16.64 &16.88 &17.34 &16.17 &17.17 &17.03\\
            &$5b_{1u}^{-1}$  &-     &17.32 &17.56 &18.03 &16.83 &17.86 &17.73\\
            &$1b_{2g}^{-1}$  &-     &18.02 &18.26 &18.65 &17.63 &18.52 &18.39\\
            &$1b_{3u}^{-1}$  &-     &19.12 &19.50 &19.67 &18.73 &19.64 &19.48\\
            &$3b_{3g}^{-1}$  &-     &19.41 &19.64 &19.87 &19.06 &19.83 &19.67\\
            &$3b_{2u}^{-1}$  &-     &20.76 &20.99 &21.17 &20.30 &21.14 &20.99\\
            &$5a_g^{-1}$     &-     &21.03 &21.32 &21.50 &20.60 &21.49 &21.36\\
  \ce{HCCCN}&$2\pi^{-1}$     &-     &11.69 &11.87 &11.41 &11.74 &11.57 &11.55\\
            &$9\sigma^{-1}$  &-     &13.44 &13.69 &13.88 &13.09 &13.82 &13.71\\
            &$1\pi^{-1}$     &-     &14.19 &14.48 &13.91 &14.15 &14.04 &13.94\\
            &$8\sigma^{-1}$  &-     &18.53 &18.81 &18.69 &18.39 &18.73 &18.72\\
            &$7\sigma^{-1}$  &-     &21.55 &21.95 &21.91 &21.35 &21.94 &21.94\\
            &$6\sigma^{-1}$  &-     &25.31 &25.89 &25.49 &25.23 &25.51 &25.50\\
  \ce{C2N2} &$1\pi_g^{-1}$    &-    &13.48 &13.71 &13.15 &13.55 &13.36 &13.32\\
            &$5\sigma_g^{-1}$ &-    &14.40 &14.66 &14.76 &14.12 &14.72 &14.66\\
            &$4\sigma_u^{-1}$ &-    &14.77 &15.04 &15.16 &14.45 &15.10 &15.03\\
            &$1\pi_u^{-1}$    &-    &15.70 &16.05 &15.37 &15.70 &15.54 &15.44\\
            &$4\sigma_g^{-1}$ &-    &23.29 &23.93 &23.44 &23.21 &23.53 &23.55\\
  \ce{C3O2} &$2\pi_u^{-1}$    &-    &10.62 &10.70 &10.32 &10.73 &10.46 &10.43\\
            &$1\pi_g^{-1}$    &-    &14.92 &15.58 &15.04 &14.21 &14.95 &14.73\\
            &$1\pi_u^{-1}$    &-    &15.72 &16.33 &16.08 &15.20 &16.02 &15.72\\
            &$5\sigma_u^{-1}$ &-    &17.10 &17.66 &18.28 &15.80 &18.01 &17.63\\
            &$6\sigma_g^{-1}$ &-    &17.28 &17.87 &18.57 &15.91 &18.23 &17.83\\
            &$4\sigma_u^{-1}$ &-    &22.22 &22.91 &22.82 &22.38 &22.99 &22.95\\
            &$5\sigma_g^{-1}$ &-    &25.88 &25.71 &26.33 &25.36 &26.73 &26.71\\
  \ce{HC4H} &$1\pi_g^{-1}$    &-    &10.20 &10.34 &9.96  &10.23 &10.09 &10.07\\
            &$1\pi_u^{-1}$    &-    &12.70 &12.94 &12.50 &12.63 &12.60 &12.52\\
            &$5\sigma_g^{-1}$ &-    &17.14 &17.39 &17.34 &16.98 &17.37 &17.35\\
            &$4\sigma_u^{-1}$ &-    &17.82 &18.10 &18.03 &17.64 &18.05 &17.99\\
            &$4\sigma_g^{-1}$ &-    &20.05 &20.41 &20.43 &19.80 &20.42 &20.39\\
            &$3\sigma_u^{-1}$ &-    &23.42 &24.05 &23.82 &23.23 &23.84 &23.81\\
            &$3\sigma_g^{-1}$ &-    &24.76 &25.38 &25.73 &24.65 &25.75 &25.76\\
    \bottomrule
  \end{longtable} 
\end{center}

\begin{center}
 \begin{threeparttable}[!htbp]
   \caption{Molecular geometry and basis sets used in vertical electron affinity (VEA)
   benchmark calculations with FCI reference values. All information is from
   ref.~\citenum{Dempwolff2021-EA}.}
   \tabcolsep=35pt
   \begin{tabular}[t]{@{}lll@{}}
   \toprule
     Molecule     & Geometry & Basis set  \\
   \midrule
    \ce{H2O}   &$R_{\ce{OH}}$ = 0.957 \AA, $\angle{\ce{HOH}} = 104.5^{\circ}$  & O:6-31+G$^*$, H:6-31++G\\
    \ce{NH3}   &$R_{\ce{NH}}$ = 1.0124\AA, $\angle{\ce{HNH}} = 106.67^{\circ}$ & N:6-31+G$^*$, H:6-31++G\\
    \ce{HF}    &$R_{\ce{HF}}$ = 0.917 \AA   &F:6-31+G$^*$, H:6-31++G\\
    \ce{CO}    &$R_{\ce{CO}}$ = 2.132 $a_0$ &cc-pVDZ \\
    \ce{HCN}   &$R_{\ce{HC}}$ = 1.064 \AA, $R_{\ce{CN}}$ = 1.156 \AA & cc-pVDZ\\
    \ce{LiH}   &$R_{\ce{LiH}}$ = 1.5957 \AA & aug-cc-pVDZ\tnote{a}\\
    \ce{CH2} & $R_{\ce{CH}}$ = 1.11656 \AA , $\angle{\ce{HCH}} = 102.4^{\circ}$  & cc-pVDZ \tnote{b} \\
    \ce{NH^{+}2} & $R_{\ce{NH}}$ = 1.0245 \AA, $\angle{\ce{HNH}} = 103.34^{\circ}$ & aug-cc-pVDZ\tnote{c}\\
    \ce{BeH^{+}} & $R_{\ce{BeH}}$ = 1.3426 \AA & aug-cc-pVDZ \\
    \ce{CH^{+}}  & $R_{\ce{CH}}$ = 1.1309 \AA & aug-cc-pVDZ \\
    \ce{CN^{+}}  & $R_{\ce{CN}}$ = 1.1718 \AA & 6-31+G \\
    \ce{NO^{+}2} & $R_{\ce{NO}}$ = 1.1934 \AA, $\angle{\ce{ONO}} = 134.1^{\circ}$ & 6-31G$$\\
    \ce{NO^{$\bullet$}2}  & $R_{\ce{NO}}$ = 1.1934 \AA, $\angle{\ce{ONO}} = 134.1^{\circ}$ & 6-31G$$ \\
    \ce{NH^{$\bullet$}2}  & $R_{\ce{NH}}$ = 1.0245 \AA, $\angle{\ce{NHN}} = 103.34^{\circ}$ & aug-cc-pVDZ\tnote{c}\\
    \ce{LiH^{$\bullet$+}} & $R_{\ce{LiH}}$ = 1.5957 \AA & aug-cc-pVDZ\tnote{a}\\
    \ce{NH3^{$\bullet$+}} & $R_{\ce{NH}}$ = 1.0124 \AA, $\angle{\ce{HNH}} = 106.67^{\circ}$ & 6-31++G$^*$ \\
    \ce{H2O^{$\bullet$+}} & $R_{\ce{OH}}$ = 0.957 \AA, $\angle{\ce{HOH}} = 104.5^{\circ}$ & 6-31++G$^*$ \\
    \ce{CO^{$\bullet$+}} &$R_{\ce{CO}}$ = 2.132 $a_0$ & cc-pVDZ \\
   \bottomrule
   \end{tabular}
   \begin{tablenotes}
   \item [a] For Li, an earlier parametrization of polarization and diffuse functions [0.00864 (s), 0.00579 (p), 0.1239 (d), 0.0725 (d)] was employed.
   \item [b] Here the cc-pVDZ basis is augmented with additional diffuse functions on C [0.015 (s)] and H [0.025 (s)] atoms.
   \item [b] A modified aug-cc-pVDZ basis without polarization diffuse functions was used on N and H atoms.
   \end{tablenotes}
 \end{threeparttable}
\end{center}

\clearpage
\begin{center}
  \begin{longtable}{cccccccc}
    \caption{35 one-particle (1p)-dominated vertical electron affinities (eV)
    for neutral molecules and cations with closed-shell reference,
    computed using FCI, EOM-EA-CCSD, EA-ADC(3), EA-ADC(4), EA-UCC3, and EA-qUCCSD.
    To avoid redundancy, the names of the approximate methods are simplified to
    CCSD, ADC(3), ADC(4), UCC3, qUCCSD in the table header.
    Target states correspond to ($N+1$)-electron
    configurations relative to the RHF reference. For \ce{CO}, \ce{HCN}, and
    \ce{NO2+}, CISDTQ results are shown in the ``FCI'' column.
    FCI and IP-ADC(3) data are from ref.\citenum{Dempwolff2021-EA}; EA-ADC(4)
    data are from ref.~\citenum{Leitner2024}.}
  \label{TS_EA1} \\
  \toprule
    \multicolumn{1}{c}{Molecule/target states} &\multicolumn{1}{c}{config.}  &
    \multicolumn{1}{c}{FCI}    &\multicolumn{1}{c}{CCSD}   &
  \multicolumn{1}{c}{ADC(3)}   &\multicolumn{1}{c}{ADC(4)} &
  \multicolumn{1}{c}{UCC3}     &\multicolumn{1}{c}{qUCCSD} \\

  \midrule
    \endfirsthead

    \multicolumn{2}{c}{\tablename\ \thetable{} -- {\sl (continued)}} \\
  \toprule
    \multicolumn{1}{c}{Molecule/target states} &\multicolumn{1}{c}{config.}  &
    \multicolumn{1}{c}{FCI}      &\multicolumn{1}{c}{CCSD}   &
  \multicolumn{1}{c}{ADC(3)}   &\multicolumn{1}{c}{ADC(4)} &
  \multicolumn{1}{c}{UCC3}     &\multicolumn{1}{c}{qUCCSD} \\
  \midrule
  \endhead

  \bottomrule
  \multicolumn{8}{r}{{\sl (continued)}} \\
  \endfoot

  \endlastfoot

  \ce{LiH}             &$^1\Sigma^+$      & & & & & & \\
    $1^2\Sigma^+$      &$3\sigma$         &0.30  &0.29  &0.31  &0.31  &0.29  &0.30  \\
    $1^2\Pi$           &$1\pi$            &-0.31 &-0.32 &-0.31 &-0.31 &-0.31 &-0.31 \\
    $2^2\Sigma^+$      &$4\sigma$         &-0.49 &-0.49 &-0.48 &-0.49 &-0.49 &-0.49 \\
    $3^2\Sigma^+$      &$5\sigma$         &-0.77 &-0.78 &-0.77 &-0.77 &-0.77 &-0.79 \\
    $2^2\Pi$           &$2\pi$            &-1.20 &-1.23 &-1.22 &-1.21 &-1.22 &-1.21 \\
    \ce{CH2}           &$^1\mathrm{A}_1$  & & & & & & \\
    $1^2\mathrm{B}_1$  &$1b_1$            &0.16  &0.08  &0.28  &0.30  &0.23  &0.22  \\
    $1^2\mathrm{A}_1$  &$4a_1$            &-0.72 &-0.74 &-0.73 &-0.73 &-0.72 &-0.72 \\
    $1^2\mathrm{B}_2$  &$2b_2$            &-1.66 &-1.65 &-1.65 &-1.65 &-1.66 &-1.66 \\
    $2^2\mathrm{A}_1$  &$5a_1$            &-1.98 &-1.98 &-1.98 &-1.98 &-2.00 &-2.00 \\
    \ce{NH3}           &$^1\mathrm{A}_1$  & & & & & & \\
    $1^2\mathrm{A}$    &$4a$              &-0.95 &-0.97 &-0.96 &-0.96 &-0.97 &-0.96 \\
    $1^2\mathrm{E}$    &$2e$              &-1.94 &-1.94 &-1.94 &-1.94 &-1.94 &-1.94 \\
    \ce{H2O}           &$^1\mathrm{A}_1$  & & & & & & \\
    $1^2\mathrm{A}_1$  &$4a_1$            &-0.94 &-0.96 &-0.94 &-0.94 &-0.95 &-0.95 \\
    $1^2\mathrm{B}_2$  &$2b_2$            &-1.88 &-1.89 &-1.88 &-1.88 &-1.88 &-1.88 \\
    $2^2\mathrm{A}_1$  &$5a_1$            &-6.27 &-6.31 &-6.27 &-6.28 &-6.30 &-6.28 \\
    $1^2\mathrm{B}_1$  &$2b_1$            &-6.68 &-6.76 &-6.71 &-6.61 &-6.75 &-6.72 \\
    $2^2\mathrm{B}_2$  &$3b_2$            &-8.25 &-8.30 &-8.27 &-8.27 &-8.28 &-8.28 \\
    \ce{HF}            &$^1\Sigma^+$      & & & & & & \\
    $1^2\Sigma^+$      &4$\sigma$         &-0.99 &-1.01 &-0.99 &-1.00 &-1.01 &-1.00 \\
    $2^2\Sigma^+$      &5$\sigma$         &-7.34 &-7.44 &-7.31 &-7.36 &-7.40 &-7.37 \\
    $3^2\Sigma^+$      &6$\sigma$         &-8.11 &-8.18 &-8.09 &-8.14 &-8.14 &-8.13 \\
    $1^2\Pi$           &2$\pi$            &-8.30 &-8.37 &-8.28 &-8.33 &-8.34 &-8.32 \\
    \ce{CO}            &$^1\Sigma^+$      & & & & & & \\
    $1^2\Pi$           &2$\pi$            &-3.62 &-3.56 &-3.53 &-3.64 &-3.56 &-3.54 \\
    $1^2\Sigma^+$      &6$\sigma$         &-9.86 &-9.79 &-9.80 &-9.77 &-9.79 &-9.79 \\
    \ce{HCN}           &$^1\Sigma^+$      & & & & & & \\
    $1^2\Sigma^+$      &6$\sigma$         &-4.41 &-4.29 &-4.48 &-4.29 &-4.48 &-4.43 \\
    $1^2\Pi$           &2$\pi$            &-4.68 &-4.33 &-4.31 &-4.37 &-4.30 &-4.30 \\
    $2^2\Sigma^+$      &7$\sigma$         &-8.90 &-8.42 &-8.53 &-8.40 &-8.49 &-8.48 \\
    \ce{NH2+}          &$^1\mathrm{A}_1$  & & & & & & \\
    $1^2\mathrm{B}_1$  &1$b_1$            &12.20 &12.24 &12.21 &12.34 &12.08 &12.08 \\
    $2^2\mathrm{A}_1$  &4$a_1$            &4.67  &4.67  &4.69  &4.69  &4.65  &4.65  \\
    $2^2\mathrm{B}_2$  &2$b_2$            &3.00  &3.02  &3.01  &3.01  &2.99  &2.99  \\
    \ce{CN+}           &$^1\Sigma^+$      & & & & & & \\
    $1^2\Sigma^+$      &5$\sigma$         &13.08 &13.43 &14.70 &14.08 &      &14.16 \\
    $2^2\Pi$           &2$\pi$            &5.23  &5.44  &6.34  &4.88  &6.67  &5.71  \\
    \ce{NO2+}          &$^1\mathrm{A}_1$  & & & & & & \\
    $1^2\mathrm{A}_1$  &6$a_1$            &10.84 &10.92 &11.24 &10.50 &11.33 &11.22 \\
    $1^2\mathrm{B}_1$  &2$b_1$            &8.02  &8.30  &8.60  &7.73  &8.77  &8.70  \\
    \ce{BeH+}          &$^1\Sigma^+$      & & & & & & \\
    $1^2\Sigma^+$      &3$\sigma$         &8.29  &8.26  &8.32  &8.31  &8.24  &8.26  \\
    $2^2\Pi$           &1$\pi$            &5.77  &5.72  &5.77  &5.77  &5.75  &5.76  \\
    $2^2\Sigma^+$      &4$\sigma$         &2.78  &2.74  &2.74  &2.78  &2.72  &2.73  \\
    $2^2\Pi$           &2$\pi$            &1.87  &1.87  &1.87  &1.93  &1.87  &1.87  \\
    $1^2\Delta$        &1$\delta$         &0.01  &-0.20 &-0.02 &0.01  &-0.03 &-0.03 \\
    \ce{CH+}           &$^1\Sigma^+$      & & & & & & \\
    $1^2\Pi$           &1$\pi$            &10.42 &10.41 &10.57 &10.52 &10.41 &10.39 \\
    $2^2\Sigma^+$      &4$\sigma$         &4.05  &4.08  &4.11  &4.10  &4.05  &4.05  \\
	\bottomrule
  \end{longtable}
\end{center}

\clearpage
\begin{center}
  \begin{longtable}{cccccccc}
    \caption{16 one-particle (1p)-dominated vertical electron affinities (eV)
    for radical, neutral triplet molecules and radical cations with open-shell
    reference, computed using FCI, EOM-EA-CCSD, EA-ADC(3), EA-ADC(4),
    EA-UCC3, and EA-qUCCSD. 
    To avoid redundancy, the names of the approximate methods are simplified to
    CCSD, ADC(3), ADC(4), UCC3, qUCCSD in the table header.
    Target states correspond to ($N+1$)-electron
    configurations relative to the UHF reference. For \ce{NO2^{$\bullet$}},
    CISDTQ results are shown in the ``FCI'' column.
    FCI and IP-ADC(3) data are from ref.\citenum{Dempwolff2021-EA}; EA-ADC(4)
    data are from ref.~\citenum{Leitner2024}.}
  \label{TS_EA2} \\
  \toprule
    \multicolumn{1}{c}{Molecule/target states} &\multicolumn{1}{c}{config.}  &
    \multicolumn{1}{c}{FCI}      &\multicolumn{1}{c}{CCSD}   &
  \multicolumn{1}{c}{ADC(3)}   &\multicolumn{1}{c}{ADC(4)} &
  \multicolumn{1}{c}{UCC3}     &\multicolumn{1}{c}{qUCCSD} \\

  \midrule
    \endfirsthead

    \multicolumn{2}{c}{\tablename\ \thetable{} -- {\sl (continued)}} \\
  \toprule
    \multicolumn{1}{c}{Molecule/target states} &\multicolumn{1}{c}{config.}  &
    \multicolumn{1}{c}{FCI}      &\multicolumn{1}{c}{CCSD}   &
    \multicolumn{1}{c}{ADC(3)}   &\multicolumn{1}{c}{ADC(4)} &
    \multicolumn{1}{c}{UCC3}     &\multicolumn{1}{c}{qUCCSD} \\
  \midrule
  \endhead

  \bottomrule
    \multicolumn{8}{r}{{\sl (continued)}} \\
  \endfoot

  \endlastfoot

  \ce{NH2^{$\bullet$}}  &$^2B_1$     & & & & & & \\
  $1^1\mathrm{A}_1$     &$1b_1$      &0.31  &0.22  &-0.08  &0.33  &-0.09  &0.01  \\
  $1^3\mathrm{B}_1$     &$4a_1$      &-0.80 &-0.81 &-0.81  &-0.80 &-0.82  &-0.81 \\
  $2^3\mathrm{B}_1$     &$5a_1$      &-4.23 &-4.26 &-4.24  &-4.23 &-4.24  &-4.23 \\
  \ce{NO2^{$\bullet$}}& $^2A_1$      & & & & & & \\
  $1^1\mathrm{A}_1$     &6$a_1$      &-0.17 &0.07  &-0.01  &-0.01 &0.19   &0.13  \\
  \ce{LiH^{*}}          &$^3\Sigma^+(2\sigma \rightarrow 3\sigma)$   & & & & & & \\
  $1^2\Sigma^+$         &$2\sigma$   &3.49  &3.21  &3.22   &3.34  &3.21   &3.22  \\
  $1^4\Pi$              &$1\pi$      &-0.01 &-0.02 &-0.01  &-0.01 &-0.02  &-0.02 \\
  \ce{NH3^{*}}          &$^3\mathrm{A}(3a \rightarrow 4a)$   & & & & & & \\
  $1^2\mathrm{A}$       &$3a$        &5.04  &4.88  &4.51   &5.13  &4.41   &4.51  \\
  \ce{H2O^{*}}          &$^3\mathrm{B}_1(1b_1 \rightarrow 4a_1)$   & & & & & &   \\
  $1^2\mathrm{A}_1$     &$1b_1$      &6.06  &5.77  &5.47   &6.07  &5.27   &5.38  \\
  $2^2\mathrm{B}_1$     &$4a_1$      &-0.15 &-0.15 &-0.08  &-0.14 &-0.13  &-0.12 \\
  $1^4\mathrm{A}_2$     &$2b_2$      &-0.77 &-0.76 &-0.73  &-0.75 &-0.77  &-0.76 \\
  \ce{LiH^{$\bullet$+}} &$^2\Sigma^+$       & & & & & & \\
  $1^1\Sigma^+$         &2$\sigma$   &7.94  &7.93  &7.94   &7.94  &7.94   &7.94  \\
  $1^3\Sigma^+$         &3$\sigma$   &4.76  &4.76  &4.75   &4.76  &4.76   &4.76  \\
  \ce{NH3^{$\bullet$+}} &$^2\mathrm{A}$     & & & & & & \\
  $1^1\mathrm{A}$       &3$a$        &10.46 &10.52 &10.14  &10.48 &10.14  &10.18 \\
  $1^3\mathrm{A}$       &4$a$        &4.46  &4.45  &4.46   &4.46  &4.45   &4.45  \\
  \ce{H2O^{$\bullet$+}} &$^2\mathrm{B}_1$   & & & & & & \\
  $1^1\mathrm{A}_1$     &$1b_1$      &12.30 &12.34 &12.04  &12.28 &11.95  &12.00 \\
  $1^3\mathrm{B}_1$     &$4a_1$      &5.30  &5.28  &5.31   &5.30  &5.28   &5.29  \\
  \bottomrule
  \end{longtable}
\end{center}

\bibliography{SI}